\documentclass{elsart3-1}

 \usepackage{graphics}
 \usepackage{graphicx}

\usepackage{amssymb}

\usepackage[english,francais]{babel}

\newtheorem{e-proposition}[theorem]{Proposition}

\newtheorem{e-definition}[theorem]{Definition\rm}

\renewcommand{\theequation}{\arabic{equation}}
\def\og{\leavevmode\raise.3ex\hbox{$\scriptscriptstyle\langle\!\langle$~}}
\def\fg{\leavevmode\raise.3ex\hbox{~$\!\scriptscriptstyle\,\rangle\!\rangle$}}

\begin{document}

\begin{frontmatter}


\selectlanguage{english}
\title{A case of strong non linearity: intermittency in highly turbulent flows }


\selectlanguage{english}
\author[label1]{Yves Pomeau},
\ead{pomeau@tournesol.lps.ens.fr}
\author[label2]{Martine Le Berre},
\ead{martine.le-berre@u-psud.fr}
\author[label3]{Thierry Lehner},
\ead{Thierry.Lehner@obspm.fr}

\address[label1]{Ladhyx, Ecole Polytechnique, 91128 Palaiseau, France}
\address[label2]{Ismo, Universit\'e Paris-Sud, 91405 Orsay Cedex, France}
\address[label3]{Luth,  Observatoire de Paris-Meudon,  92195 Meudon, France}

\medskip
\begin{center}
{\small Received *****; accepted after revision +++++\\
Presented by Â£Â£Â£Â£Â£}
\end{center}

\begin{abstract}
It has long been suspected that flows of incompressible fluids at large or infinite Reynolds number (namely at small or zero viscosity) may present finite time singularities. We review briefly the theoretical situation on this point. We discuss the effect of a small viscosity on the self-similar solution of the Euler equations for inviscid fluids. Then we show that single point records of velocity fluctuations in the Modane wind tunnel display correlations between large velocities and large accelerations in full agreement with scaling laws derived from Leray's equations (1934) for self-similar singular solutions of the fluid equations. Conversely those experimental velocity-acceleration correlations are contradictory to the Kolmogorov scaling laws.   

{\it To cite this article: Y. Pomeau, M. Le Berre and T. Lehner, C. R.
Mecanique -- (2018).}

\vskip 0.5\baselineskip

\selectlanguage{francais}
\noindent{\bf R\'esum\'e}
\vskip 0.5\baselineskip
\noindent
{\bf Un cas de forte nonlin\'earit\'e: l'intermittence en milieu turbulent \`{a} grand nombre de Reynolds}.
On pense depuis longtemps que les \'ecoulements fluides incompressibles \`a grand, sinon infini, nombre de Reynolds pr\'esentent des singularit\'es localis\'ees en temps et en espace. Nous \'etudions l'effet d'une petite viscosit\'e sur les solutions auto-semblables des \'equations des fluides.  Nous montrons ensuite que des enregistrements de fluctuations de vitesse dans la soufflerie de Modane pr\'esentent des corr\'elations entre grandes vitesses et grandes acc\'el\'erations en accord complet avec les lois d'\'{e}chelle d\'{e}duites des  solutions  auto-similaires des \'equations trouv\'ees par Leray en 1934. En revanche ces corr\'elations sont en contradiction avec les lois d'\'{e}chelle d\'eduites de la th\'eorie de Kolmogorov. 

{\it Pour citer cet article~: Y. Pomeau, M. Le Berre and T. Lehner, C. R.
Mecanique ---(2018).}

\keyword{Navier-Stokes; Euler; Turbulence; Singularities; Leray; Intermittency }
\vskip 0.5\baselineskip
\noindent{\small{\it Mots-cl\'es~:} Navier-Stokes; Euler; Turbulence; Singularit\'es; Leray; Intermittence }}
\end{abstract}
\end{frontmatter}


\section{Foreword}
Over the years Pierre Coullet developed an outstanding research devoted to many aspects of nonlinearity in Science. With his very sure taste he chose topics with a deep  geometrical underpinning, non linearity being only one element in the structure of the scientific question. In Fluid mechanics nonlinearity and geometry concur to bring forward difficult and fascinating questions. We think first to the transition to turbulence by cascade of period doubling, predicted  almost simultaneously by Pierre Coullet and Charles Tresser  \cite{crasCTr} and by Mitch Feigenbaum \cite{mitch}. This scenario of transition was observed slightly afterwards in fluid experiments at Ecole normale laboratory by Jean Maurer and Albert Libchaber \cite{maurer}. The understanding of the transition to turbulence with a few degrees of freedom did not end research in fluid turbulence, a difficult field where real progress has always been very slow. The next step was the realization that the transition to turbulence in large systems with many degrees of freedom  belongs to the class of directed percolation \cite{directedperc}, but this cannot be seen as the end the story. A big open question remaining in fluid turbulence was raised in the 1949 paper by Batchelor and Townsend \cite{batchT} where the authors discuss observations of large velocity fluctuations they attribute in, we believe, a not fully convincing way to the large wavenumber limit of the Kolmogorov cascade  \cite{K41}. This assumption (like many, if not most, works in turbulence theory) bypass any discussion of the time dependence of the fluctuations of the turbulent flow and their link to the basic fluid equations, whereas the observed intense and short bursts of turbulence have obviously something to do with the time dependence of solutions of the fluid equations, a point we expand on below.  We hope this contribution will show our admiration for Pierre and will be also of interest for fluid mechanics. 

\section{Introduction}
\label{intro}

 One outstanding problems of turbulence in fluids was posed by Batchelor and Townsend in 1949 \cite{batchT} and can be stated as follows. Kolmogorov theory predicts a spectrum of velocity fluctuations decaying like $k^{-5/3}$ at large wave numbers, the Kolmogorov-Obukhov spectrum. Measurements made over the years agree well with this prediction  \cite{expmod}-\cite{exp2mod}. Therefore it was somewhat surprising to observe also that the largest velocity and acceleration fluctuations in a turbulent flow are short lived and are also associated with short distances. This looks  contradictory with the Kolmogorov-Obukhov spectrum which predicts that the intensity of the fluctuations  of velocity decreases as the length scale (the inverse wave number) decreases, because the statistical weight of large velocities  fluctuations (and acceleration)  is not small,
particularly in  the Modane experiment,  where  about $4$ per cent of the recorded
data are
for acceleration larger than $2.5$ in units of its standard deviation.  This phenomenon is called intermittency. A particular consequence of
these  very intense and quick  bursts observed  in the record,  is the
wings widening  of the probability distribution of the acceleration,
a property which cannot be explained in a theory with a
single scaling parameter, as  the one in the original Kolmogorov theory
of 1941 \cite{K41}.

To describe fluid motion, besides Kolmogorov or Kolmogorov-inspired statistical theories, there are basic equations, the Navier-Stokes (NS) equations becoming the Euler equations in the inviscid limit. It makes sense to come back to those fundamental equations to see if the phenomenon of intermittency is explainable by them and, in particular, if predictions could be made concerning it. This is the purpose of this paper, which includes an analysis of velocity data recorded in the big wind tunnel of Modane in Southern France. 
 
As no general solution of either the NS or Euler equations is known, it could seem hopeless to base a theory on solutions of those equations. However the situation is not as bad as one could believe first, because of the idea of self-similar solutions for the fluid equations, an idea going back to Leray \cite{leray}.
We explain in  Sec.\ref{sec:sssol}-\ref{sec:Euler-Leray} what are those self-similar solutions. In  Sec.\ref{sec:smallvisc} we  apply this to predict the occurrence of  quasi-singularities, namely singular solutions of the Euler equations becoming smooth under the effect of viscosity. This relies on two assumptions, first that the Euler equations have a finite time singularity whereas the NS (Navier-Stokes) equations have not. Based on this we predict a relation between the large fluctuations of the velocity and of the acceleration which is  amazingly well verified by hot-wire records made in  Modane's wind tunnel, Sec. \ref{sec:vo}-\ref{sec:modane}.

\section{Self-similar fluid equations}
\label{sec:sssol}

In 1934 Leray \cite{leray} published a paper on the Navier-Stokes equations  for an incompressible fluid in 3D,
\begin{equation}
\frac{\partial{\bf{u}}}{\partial t} + {\bf{u}}\cdot \nabla {\bf{u}} = - \nabla p  + \nu \nabla^2 {\bf{u}}   \qquad \nabla \cdot {\bf{u}} = 0 
\textrm{,}
\label{eq:NS1}
\end{equation}
where ${\bf{u}}$ is the velocity field, $p$  is the pressure and  $\nu$  is the kinematic viscosity of the fluid. Without the viscosity term, one obtains the  Euler equations (see next section). 
In his paper Leray introduced many important ideas, among them the notion of weak solution and also the problem of the existence (or not) of a solution becoming singular after a finite time, when starting from smooth initial data.
He looked for solutions of the self-similar type, 
\begin{equation}
{\bf{u}}( {\bf{r}}, t) = (t^*- t)^{-\alpha}  {\bf{U}} ( {\bf{r}}(t^*- t)^{-\beta})  \qquad 
 p( {\bf{r}}, t) = (t^*- t)^{- 2\alpha} P ( {\bf{r}}(t^*- t)^{-\beta})
\textrm{,} 
\label{eq:self}
\end{equation}
 where $t^*$ is the time of the singularity (set to zero later),  $\alpha$ and $\beta$ are real positive exponents to be found and the pair of functions $({\bf{U}},P)$ with upper-case letters  is to be derived from Euler, or NS equations, see below. 
That such a velocity field is a solution of Euler or NS equations implies to balance the two terms on the left hand side of (\ref{eq:NS1}), which behave respectively as $t^{-(\alpha+1)}$  and $t^{-(2\alpha+\beta)}$.  It yields a first relation between the two parameters,
\begin{equation}
  \alpha + \beta =1
 \textrm{.}
 \label{eq:alpha1}
\end{equation}
and the re-scaled equation for ${\bf U}$\begin{equation}
(\alpha {\bf{U}}  + \beta {\bf{R}}\cdot \nabla {\bf{U}}) + {\bf{U}}\cdot \nabla {\bf{U}} = - \nabla P  \qquad   \nabla \cdot {\bf{U}} = 0
\textrm{.} 
 \label{eq:Leray-ab}
\end{equation}

In the case of Navier-Stokes equation, the balance with the dissipative term  $\nu \nabla^2  {\bf{u}}$, of order  $t^{-(\alpha+2\beta)}$, imposes $\beta=1/2$,   which yields  the exponents found by Leray,
\begin{equation}
 \alpha = \beta =1/2
 \textrm{.}
 \label{eq:alpha2}
\end{equation}
Let us give an outlook of the derivation of Leray's  equation  for $({\bf{U}},P)$ (see for example \cite{giga} but not done in this way by Leray). We consider the case $t-t^{*} < 0$   leading to what is sometimes called backward self-similar equation.  If the NS equations admits self-similar solutions, the set $({\bf{u}},p)$  must be of the form,
 \begin{equation}
{\bf{u}}( {\bf{r}}, t) = \sqrt\frac{\nu}{-t}  
 {\bf{U}} \left( {\bf{r}}(- \nu t)^{-\frac{1}{2}}\right)  \qquad p( {\bf{r}}, t) = \left(\frac{\nu}{-t}  \right) P\left( {\bf{r}}(- \nu t)^{-\frac{1}{2}} \right)
\textrm{.} 
\label{eq:self2}
\end{equation}
where  $t$ is  for $t-t^{*}$ and $\nu$ is the kinematic viscosity. 

Introducing the logarithmic time $\tau= - \log(- t)$,  plus additional  changes of variables ${\bf{R}}={\bf{r}}(- \nu t)^{-1/2}$,  ${\bf{U}}({\bf{R}},\tau) = \sqrt\frac{-t} {\nu}  {\bf{u}}( {\bf{r}}, t)$,  $P({\bf{R}},\tau)=  \left(\frac{-t}{\nu}  \right)  p( {\bf{r}}, t)$,
the equations for the pair of functions $({\bf{U}}({\bf{R}},\tau),P({\bf{R}},\tau))$  become,
\begin{equation}
\frac{\partial{\bf{U}}}{\partial \tau} + \frac{1}{2}( {\bf{U}} + {\bf{R}}\cdot \nabla {\bf{U}}) + {\bf{U}}\cdot \nabla {\bf{U}}= - \nabla P  +  \nabla^2 {\bf{U}}   \qquad \nabla\cdot  {\bf{U}}=0
\textrm{.}
\label{eq:NS-Giga1}
\end{equation}

The interest of using such a time dependence is that, for  original differential equations of first order with respect to  time $t$, the new differential equation is still first order and autonomous with respect to $\tau$. 
The pair $({\bf{u}},p)$ is a self-similar solution of NS or Euler equations if and only if ${\bf{U}} ={\bf{U}}({\bf{R}})$ and $P=P({\bf{R}})$ are fixed points depending only of ${\bf{R}}$, that gives the Leray equation,
\begin{equation}
 \frac{1}{2}( {\bf{U}} + {\bf{R}}\cdot \nabla {\bf{U}}) + {\bf{U}}\cdot \nabla {\bf{U}}= - \nabla P  +  \nabla^2 {\bf{U}}   \qquad \nabla\cdot  {\bf{U}}=0
\textrm{.}
\label{eq:NS-Leray}
\end{equation}
 In the following, equations of the form (\ref{eq:NS-Leray})  are called NS-Leray equations if the viscosity is non-zero and Euler-Leray equations whenever the $  \nabla^2 {\bf{U}} $ in  (\ref{eq:NS-Leray})  is absent and if, with respect to the scaled variables (\ref{eq:self2}) the circulation, $\Gamma$, carried by the flow towards the singularity replaces the kinematic viscosity $\nu$.

Over the years the search for solutions of (\ref{eq:NS-Leray}) motivated many works, mostly by mathematicians. The main effort was to try to prove (or disprove) the existence of such singularities assuming properties of the initial data \cite{rieusset}. Other attempts have been directed toward a direct numerical solution of NS and/or  Euler equations, with the purpose of showing they have or not a finite time singularity \cite{gibbon}. 

\section{ Euler-Leray equations}
\label{sec:Euler-Leray}

 In the case of Euler equations, the existence of a self-similar solution imposes (\ref{eq:alpha1}), but the balance condition with the dissipative term $\nu \nabla^2  {\bf{u}}$ does not hold, allowing others sets of exponents different from (\ref{eq:alpha2}).
One exponent,  $\beta$ for instance, is seemingly free, namely it does not follow from simple algebraic manipulation of the Euler equations. There are several possibilities to get a second relation between the two exponents $\beta$ and $\alpha$. This relies on the existence of conservation laws and the final result depends on what conservation law is considered. 

Let consider first the conservation of circulation on closed curves.  The circulation $\Gamma$ along a closed curve carried by the flow toward  the singularity, is of order $t^{\beta-\alpha}$. Therefore the conservation of circulation implies $ \alpha = \beta$, that gives (\ref{eq:alpha2}), namely the  same exponents as for  the Navier-Stokes case.  
Moreover the velocity scales like $u(r, t) \sim   \sqrt\frac{\Gamma}{-t}  $ near the singularity. With such a choice, the  total energy of solutions of the self-similar problem is diverging, but this divergence of the energy  does not imply the absence of singularity of finite energy of  a different type. 

For reasons explained in  \cite{modane} we shall consider the exponents (\ref{eq:alpha2}) ensuring conservation of circulation. This yields self-similar solutions like (\ref{eq:self2}), and Euler-Leray  first equation given by (\ref{eq:NS-Leray})  without the viscosity term,or
\begin{equation}
\frac{\partial{\bf{U}}}{\partial \tau} + \frac{1}{2}({\bf{U}}  + {\bf{R}}\cdot \nabla {\bf{U}}) + {\bf{U}}\cdot \nabla {\bf{U}} = - \nabla P  \qquad   \nabla \cdot {\bf{U}} = 0
\textrm{.} 
 \label{eq:Euler-Leray1}
\end{equation}

 Now let us consider more generally, a self-similar solution of Euler equations of the form (\ref{eq:self}) with arbitrary exponents $\alpha,\beta$ (see eqn. (\ref{eq:Leray-ab})). If one considers instead of the conservation of circulation, the conservation of  total energy in the collapsing domain, one must satisfy the constraint $- 2 \alpha + 3 \beta = 0$, together  with (\ref{eq:alpha1}), that yields in the inviscid case
\begin{equation}
\alpha = 3/5 \qquad  \beta = 2/5
\textrm{,} 
 \label{eq:ab}
\end{equation}
which are the Sedov-Taylor  exponents  \cite{sedovtaylor}. 
No set of singularity exponents can satisfy both constraints of energy conservation {\it{and}} constant circulation on carried closed curves. In the following we mainly  focus on the case originally proposed by Leray, $\alpha=\beta=1/2$, associated to the constraint of circulation conservation, although the case of Sedov-Taylor exponents (\ref{eq:Leray-ab}) is potentially promising, as illustrated by Fig.\ref{fig:Leray}-(b), and could be investigated by using the same approach as the one proposed below.

Of course it is highly desirable to have a non trivial solution of equation  (\ref{eq:Euler-Leray1}) either in an analytic form or resulting from numerical analysis. This may include an explicit dependence with respect to the ``time" $\tau$, and  does not seem to be as hopeless as one might think first.  Let us outline a possible solution of this problem.
 The idea is to consider a solution of equation (\ref{eq:Euler-Leray1}) which is close to a steady solution of Euler equation in the axisymmetric case. This  could be valid for large amplitude solutions, and amounts to solve, at leading order,  the  non linear part of equation (\ref{eq:Euler-Leray1}), 
\begin{equation} 
  {\bf{U}}^{0}\cdot \nabla {\bf{U}}^{0} = - \nabla P^{0}  \qquad   \nabla \cdot {\bf{U}}^{0} = 0  \textrm{.}
 \label{eq:Euler-stat}
\end{equation}
 In the limit of large $U$,
the two other terms, namely 
 $\frac{\partial{\bf{U}}}{\partial \tau} + \left(  \frac{3}{5} {\bf{U}}  +  \frac{2}{5} {\bf{R}}\cdot \nabla {\bf{U}}\right)$ 
are relatively small perturbations.  The expansion near the leading order solution $ {\bf{U}}^{0}$ leads to two solvability conditions which are satisfied by tuning the amplitudes of two modes, $A_{1,2} \exp(i\omega_{1,2}\tau)$ oscillating in  time $\tau$ near the steady state solution. Such oscillations could manifest themselves in the time records, as observed in experimental signals of turbulent flows, see Fig.\ref{fig:burst} which displays  a decreasing oscillatory behavior in the decaying phase of huge fluctuations. In the case of the Euler equations there is no direction of time because of the symmetry of the equations under time reversal, so that oscillations after the singularity may have the same explanation as oscillations before the singularity, although their amplitude is different of the ones before the ``singularity" because of the increased viscous dissipation expected near the time of the singularity.

To conclude on the Euler-Leray equations, they yield a well defined schema for the existence of solutions of the Euler equations in 3D becoming singular in a finite time and at a single point. A by-product of this analysis is the set of exponents of the singularity which may be compared to experimental data for the big fluctuations observed in the time-records of the velocity in a turbulent flow, as done below.

One motivation for working on Euler-Leray singularities is their possible connection with the phenomenon of intermittency in high Reynolds number flows, a point we have not found in the literature, although numerous works are devoted to a direct investigation of Euler equations  (see the impressive list in \cite{gibbon}). This possible relevance of Euler-Leray singularities for explaining observed features of  turbulent flows raises several questions. Among them one may quote the following: 

\begin{enumerate}
\item What is the difference between Euler-Leray and  NS-Leray singularities? 

\item  What is specific to our interpretation in terms of Leray singularities compared to other schema for intermittency? 

\item What would be specific of  Euler-Leray singularities in a time record of large Reynolds number flow? 

\end{enumerate}
We comment about previous points below:

\begin{enumerate}
\item 
Little is known about this difference, in particular do both have nontrivial solutions, or does none has nontrivial solutions or only one has nontrivial solution? Mathematicians have obtained over the years various constrains on the functional space where such solutions could exist.  This point (i) is discussed below in Section \ref{sec:smallvisc} devoted to the understanding of the effect of adding a small but finite viscosity to the singularities of the Euler-Leray equations in order to agree with the real physical situation of viscous fluids. 

\item
If intermittency is linked to Leray-like singularities, they yield automatically a strong correlation between large values of the velocity and of the acceleration (see below). Compared to predictions derived from Kolmogorov theory this correlation is a strong indication of the occurrence of Leray-like singularities near large fluctuations.  It is fair to say however that, as far as we know, Kolmogorov himself has never mentioned this question of finite time singularity of either NS or Euler equations. So it would be unfair to attribute to him any statement about those singularities. 

\item
Both in Euler-Leray and NS-Leray ($\alpha=\beta=1/2$)  the velocity field at the time of blow-up scales like $1/r$, $r$ distance to the singularity, so that a knowledge of the flow structure near a singularity would not help to distinguish between the two kinds of singularity. 
\end{enumerate}

Our analysis of the experimental data relies on a relationship between velocity and acceleration at the same time (given by the time records of velocity). The theory for this relationship is fairly simple.
First the velocity scales like $u(r, t) \sim (-t)^{-1/2} \Gamma^{1/2}$ near the singularity, as written above. 
In the absence of viscosity  the order of magnitude of the circulation  $\Gamma$ is constant in the collapsing domain, and the typical Reynolds number of the small length scales does not tend to zero, but stays constant because the velocity grows at the same pace as the space scale decreases.  The circulation is the one along a closed curve  which is wholly carried by the flow in the collapsing domain. A discussion of the existence of such closed curve in the case of solution of Euler-Leray close to a steady solution of Euler equations will be given soon \cite{future}.
These properties are opposite to what is expected from standard ideas on turbulence according to which the role of viscosity increases and the Reynolds number decreases  as one approaches small spatial scales.

From the scaling laws of the velocity one derives immediately the one for the acceleration $\gamma(r, t)$. This acceleration is not the one of a particle carried by the flow, sometimes called Lagrangian acceleration, but only the time derivative of the fluid velocity measured at a given point, the quantity we have access to from hot wire measurements, also named Eulerian acceleration. This acceleration scales like $\gamma(r, t) \sim (-t)^{-3/2} \Gamma^{1/2}$. Accordingly one finds the time independent relation,  
\begin{equation}
u^3  \sim  \gamma \Gamma  \qquad  \textrm{for (\ref{eq:alpha2})} 
\textrm{.} 
 \label{eq:cubv}
\end{equation}
Let us make a step aside to see how the latter relation changes  if one assumes  that  the energy $E$  is conserved in the singular domain.  Using the definition of the self-similar solution for the velocity field  of type (\ref{eq:self}) with Sedov-Taylor exponents given by (\ref{eq:ab}), we get $u(t) \sim (E/ (-t)^{3})^{1/5}$. In that case the time independent relation between $u$ and $\gamma \sim u/(-t)$ becomes
\begin{equation}
u^{8/3}  \sim  \gamma E^{1/3}  \qquad  \textrm{for (\ref{eq:ab})} 
 \textrm{.} 
 \label{eq:sedov}
\end{equation}
The two relations (\ref{eq:cubv}) and  (\ref{eq:sedov}) predict that large accelerations are associated to large velocity fluctuations, and  should display a similar power-law dependence $\gamma \propto u^{z}$ with $z\simeq 3$.  We shall return to this point in section \ref{sec:modane}.
 
Let us now turn to the relationship between velocity and acceleration derived from the Kolmogorov scaling. The starting point is the Kolmogorov relation $ u_r  \sim  (\epsilon r)^{1/3}$ where $u_r$ is the typical change of velocity over a distance $r$ and $\epsilon$ the  rate of dissipation of the kinetic energy  density per unit mass of the turbulent fluid. With those scaling the time derivative  of the velocity  is of order $\gamma  \sim \epsilon^{2/3} r^{-1/3}$.
 Therefore  one has the following relationship, independent on $r$, between $u_r$ and $\gamma$,  
\begin{equation} 
u_r \gamma \sim  \epsilon 
 \textrm{,} 
 \label{eq:Kolm}
\end{equation}
an expression which can be derived directly from the definition of $\epsilon$. 

Note that if the Taylor hypothesis is used,  the partial time derivative of the  velocity should be equal to $v_{0}\partial u /\partial x$, where $v_{0}$ is the  advection velocity. 
In the case of a self-similar solution  like (\ref{eq:self2}) the   Eulerian acceleration becomes $\gamma_{\rm Taylor} \sim v_{0} U'/t$ and  (\ref{eq:cubv}) becomes
\begin{equation}
u_r^{2} \sim  \frac{\Gamma}{v_{o}}\gamma_{\rm Taylor}  \textrm{.} 
 \label{eq:cubvb}
\end{equation}
On the other hand Kolmogorov scalings lead to
$\gamma_{\rm Taylor} \sim v_{0}\epsilon u_{r}^{-2}$, and the  relation (\ref{eq:Kolm}) between $u_{r}$ and $\gamma_{\rm Taylor} $ must be replaced by
\begin{equation}
u_r^{2} \gamma_{\rm Taylor} \sim  v_{o}\epsilon  \textrm{,} 
 \label{eq:Kolmb}
\end{equation}

The two  relations (\ref{eq:cubv})-(\ref{eq:Kolm}) deduced without the Taylor hypothesis  (and also the two relations (\ref{eq:cubvb})-(\ref{eq:Kolmb}) deduced for the case of  frozen turbulence)
are so sharply  different that it makes sense to  see wether some of them agree  with experimental data, as done in section \ref{sec:modane}. We discuss in Sec.~\ref{sec:modane2} the pertinence of using Eulerian acceleration $\gamma(r, t)=\partial{u}/\partial{t}$ to test those scaling laws.

 \section{Effect of a small viscosity on singularities of the Euler-Leray equations}
 \label{sec:smallvisc}
 This section does not rely on proved results on solutions of NS-Leray or Euler-Leray equations. It attempts to show a possible scenario of what happens concerning the occurrence of singularities in the physically relevant situation of a large, but not infinite, Reynolds number. Our main assumption  is that Euler-Leray has a \textit{bona fide} solution whereas NS-Leray has not. This statement is  unproved on either side as far as we can tell and requires some explanation. 
 
Over the years mathematicians studied rather intensively Leray equation~\cite{rieusset}. To our knowledge a still incomplete understanding has been reached yet. In the case of NS-Leray equation,  various negative results have been presented, which exclude (non-zero) solutions belonging to certain functional space.  We think that  the expected slow decay like $1/r$ of the solution of NS-Leray is  a source of difficulties to reach  a definite conclusion. Nevertheless we shall make the hypothesis 
of the absence of solutions of NS-Leray with this long range dependence. Obviously, we exclude unbounded solutions at large distances. The case of Euler-Leray seems to be more complex. At this point, as far as we call tell, the situation is very uncertain. Some numerical simulations point to a self-similar solution with measurable exponents whereas mathematics exclude the existence of  such solutions or give bounds, including lower bounds, for norm of solutions depending on a free exponent introduced at the beginning. This seems not to exclude a non-trivial solution. 

 Below we shall assume that:

\begin{enumerate}
\item  There is no convenient non-zero solution of NS-Leray. By ``convenient" we mean that ${\bf{U}}({\bf{R}}) $ is a smooth solution (not growing at infinity), in other words ${\bf{u}}({\bf r},t=0) $ is non singular. 
 
\item Euler-Leray has convenient non-zero solutions.  

\end{enumerate}

 The next step in our analysis is to consider the NS-Leray singularity problem in the (realistic) limit of a small but non vanishing viscosity. It is legitimate to study the behavior of an initial condition that is exactly the solution of Euler-Leray equations and to find what happens to this solution  if the viscosity is small but not zero. Within our assumption of lack of solution of NS-Leray, the evolution of such an initial condition is changed dramatically by a small viscous term as we are going to explain. 
 
 The Euler-Leray equations have an interesting structure, pointed out in  \cite{YP}, they are invariant under dilation. This means that if  (\ref{eq:Euler-Leray1}) have a solution  ${\bf{U}}({\bf{R}})$, then the function $\mu {\bf{U}}( \mu{\bf{R}})$ is also a solution, with $ \mu $ arbitrary real number. This continuous symmetry in the set of solutions of the Euler-Leray equations will play, as usual in this type of situation, an important role in the  perturbation brought by a small change in the equations. Such a change is the addition of a small viscosity, which breaks the dilation invariance because NS-Leray equations are not invariant under dilation at constant non vanishing viscosity, unless $ \mu = -1$, which does not correspond to a continuous symmetry. 
 Note that besides this dilation symmetry there is also a continuous symmetry under rotations which is preserved by the viscosity term and so does not bring any dynamics of the parameter $\mu$, contrary to the breaking of dilation invariance. 
 
Because of the breaking of dilation invariance, it is not possible to find by regular expansion a solution of NS-Leray equations 
 close to a solution of Euler-Leray equations in the limit of a small but non-zero viscosity. This is because at first order with respect to the small viscosity one finds a solvability condition which is impossible to satisfy in the framework of the equations of similarity as they stand. 
 Let us sketch a more detailed explanation.  To get a solution of (\ref{eq:NS-Giga1}) for small viscosity (more properly $\frac{1}{Re} = \nu/\Gamma \ll 1$ , we start from a solution $ \overline{{\bf{U}}}({\bf{R}})$  of equation (\ref{eq:Euler-Leray1}), associated to an arbitrary value $ \mu = 1$ of the dilation parameter.  The solution ${\bf{U}}_{EL} = \mu \overline{{\bf{U}}}( \mu{\bf{R}})$  of (\ref{eq:Euler-Leray1}) is  the leading order term of our unknown solution of (\ref{eq:NS-Giga1}), expanded in powers of  the small parameter $\nu$.
 The effect of a non vanishing viscosity is to cause a drift of the solution of Euler-Leray equations in the space of the parameter $ \mu $, and also to introduce another time dependence of the singular solutions, via the logarithm $\tau = -\ln(-t) $, $ t = 0$ being the instant of the blow-up. 
With respect to this new ``time" variable the blow-up time is sent to plus infinity. 

 At first order with respect of the small parameter $\nu$, a small (unknown) perturbation  $ {{\bf{U}}}_c ( {\bf{R}}, \mu)$ added to $ {\bf{U}}_{EL} ( {\bf{R}}, \mu)$ must satisfy
 a solvability condition deduced 
  by introducing a slow dependence of $\mu$ with respect to $\tau$, hence a slow variation with respect to $\tau$ of the solution.
The first correction ${\bf{U}}_{c}$ is of order $ \nu/ \Gamma $. It has to satisfy a linear equation derived by putting  ${\bf{U}}_{EL}  + {\bf{U}}_{c}$ into equation (\ref{eq:NS-Giga1}) and keeping only terms linear with respect to ${\bf{U}}_{c}$, $\nu$ and $ \partial /\partial{\tau}$. We get,
\begin{equation}
\frac{ \partial {\bf{U}}_{EL}}{\partial{\tau}} + {\mathcal{L}}  [ {\bf{U}}_{EL}]  {\bf{U}}_{c} = \nu \nabla^2 {\bf{U}}_{EL} 
\textrm{,} 
 \label{eq:self2.5}
\end{equation}
where ${\mathcal{L}}$ is a linear operator acting on functions of ${\bf{R}}$, derived by  linearization of (\ref{eq:Euler-Leray1}) near the solution ${\bf{U}}_{EL}$. This operator is such that 
${\mathcal{L}}  [ {\bf{U}}_{EL}] {\bf{U}}_{d}=  0 \textrm{,} $ for $ {\bf{U}}_{d}  =  \frac{ \partial {\bf{U}}}{\partial{\mu}} $ because of the dilation invariance of the Euler-Leray equations. In technical terms the function $  \frac{ \partial {\bf{U}}}{\partial{\mu}}$ belongs to the non-empty kernel of the linear operator ${\mathcal{L}}  [{\bf{U}}_{EL}]$ (we set aside for the moment the question of the way the pressure enters into this). Define now an inner product, a real number in the space of functions of ${\bf{R}}$, namely a bilinear quantity $\left<  {\bf{U}}_{e}( {\bf{R}}) |  {\bf{U}}_{f}( {\bf{R}}) \right>$ where $e$ and $f$ are arbitrary indices. This inner product must be defined by convergent integrals, which requires some care because many functions under consideration decay slowly at large $R$.   

The dynamical equation for $\mu (\tau)$ is derived as a solvability condition for equation (\ref{eq:self2.5}), because once the equation is multiplied by the kernel of the operator adjoint of  ${\mathcal{L}}$, the unknown function ${\bf{U}}_{c}$ disappears completely out of equation (\ref{eq:self2.5}) and the only freedom to cancel the result is to impose an equation of motion for $\mu (\tau)$. This is done by writing $\frac{ \partial {\bf{U}}_{EL}}{\partial{\tau}}  =   \frac{ \mathrm{d}\mu}{\mathrm{d}\tau} \frac{ \partial {\bf{U}}_{EL}}{\partial{\mu}}$, with the final result,
\begin{equation}
\frac{ \mathrm{d}\mu}{\mathrm{d}\tau} \left<U^{ \dagger} ({\bf{R}}) | \frac{\partial {\bf{U}}_{EL}}{\partial{\mu}}\right>  = \nu  \left<U^{ \dagger} ({\bf{R}}) |\nabla^2 {\bf{U}}_{EL}\right> 
\textrm{,} 
 \label{eq:self2.6}
 \end{equation}
 In this equation, the function $U^{ \dagger} ({\bf{R}}) $ belongs to the kernel of the linear operator conjugate of the kernel of ${\mathcal{L}}  [{\bf{U}}_{EL}]$ with the inner product still to be chosen, that is  ${\mathcal{L}} ^\dagger U^{ \dagger} ({\bf{R}}) =0$. The end result of this is a dynamical equation for the dilation parameter $\mu$. 
 
 Let us turn now to the definition of the inner product $\left<  {\bf{U}}_{e}( {\bf{R}}) | {\bf{U}}_{f}( {\bf{R}})\right>$. This is in principle arbitrary, except that a change in its definition leads to a change in the operator conjugate of ${\mathcal{L}}$ and then of the function $U^{ \dagger} ({\bf{R}}) $. The velocity field ${\bf{U}}_{EL}$ decays like $1/R$ at large $R$. The same kind of argument used to derive the long distance behavior of ${\bf{U}}_{EL}$ shows also that $U^{ \dagger} ({\bf{R}}) $ decays like $1/R$ at large $R$. Let us introduce the usual inner product as the integral over space of the scalar product of two vector fields, 
 \begin{equation}
\left<  {\bf{U}}_{e}( {\bf{R}}) | {\bf{U}}_{f}( {\bf{R}})\right> = \int {{\bf{U}}}_{e}( {\bf{R}}) \cdot  {\bf{U}}_{f}( {\bf{R}}) \, \mathrm{d}{ {\bf{R}}}
\textrm{.} 
 \label{eq:self2.7}
 \end{equation}
It is not hard to check that $\nabla^2 {\bf{U}}_{EL}$ decays like $1/R^3$ as $R$ becomes large, whereas $U^{ \dagger} ({\bf{R}})$ is of order $1/R$ in the same limit. Therefore the integrand on the right-hand side of the solvability condition (\ref{eq:self2.6}) decays like $1/R^4$ at $R$ large, so that the integral converges at large distances. The left-hand-side is less simple. The field $U^{ \dagger} ({\bf{R}}) $ decays like $1/R$ at large $R$.  The derivative $\frac{\partial {\bf{U}}_{EL}}{\partial{\mu}}$ does not include terms of order $1/R$ because the term $1/R$  of  $\mu {\bf{U}}_{EL}(\mu{\bf{R}})$  is  independent of $\mu$. Therefore the first non zero contribution at $R$ large to this derivative is of order $1/R^3$, so that the inner product  $ \left<U^{ \dagger} ({\bf{R}}) | \frac{\partial {\bf{U}}_{EL}}{\partial{\mu}}\right> $ is given by a converging integral at $R$ large. 

There remains to settle the question of the pressure. This can be done, at least formally, by relating the pressure to the square of the velocity field by taking the divergence 
of (\ref{eq:Euler-Leray1}). This yields a Poisson equation for the pressure where the source term is the gradient of Reynolds tensor. This Poisson equation can be solved formally for the pressure. The result can be inserted into the equation for ${\bf{U}}$ which becomes an equation without the pressure but with an integral term. This allows to use the formalism introduced above and yields an expression for $\frac{\mathrm{d}\mu}{\mathrm{d}\tau}$ that is explicit, but rather complicated. The integrals giving the inner products are still converging because, as was shown in  \cite{YP}, the pressure decays like $1/R^3$ at large $R$. 

Because we have no explicit form, either analytical or even numerical of the field ${\bf{U}}_{EL}$ solution of Euler-Leray equations, it is not possible to say anything precise concerning the solution of equation  (\ref{eq:self2.6}), namely concerning the ultimate fate of the self-similar solution of Euler-Leray  once the viscosity is turned on. This equation for $\mu( \tau)$ has a very simple mathematical structure, being first order with respect to $\tau$ and autonomous. Even without knowing explicitly ${\bf{U}}_{EL}$ one can say that there are two possibilities: either the solution $\mu( \tau)$ tends to zero as $\tau$ tends to infinity or tends to a non zero fixed point. In the first case the approximation made in deriving equation  (\ref{eq:self2.6}) breaks down at a certain time $ \tau$ because, if $\mu$ tends to zero, the coefficient of the right-hand side which should remain small by assumption, is of order $\nu/(\Gamma \mu^2)$, $\Gamma$ being the initial value of the circulation. The quantity $\nu/(\Gamma \mu^2)$, which is initially small because $\nu\ll \Gamma$, grows indefinitely as $\mu$ tends to zero. Therefore the initial assumption of a small viscosity breaks down as $\tau$ tends to infinity, namely as time gets close to the singularity time. This means that a new regime is reached where viscosity cannot be considered anymore as small. It is reasonable to guess that in the absence of external forcing, the solution decays to zero then. Of course in a real turbulent flow there is always forcing by fluctuations of pressure so that the time dependence does not stop at this time and continues. This could be represented mathematically by random forcing term in the fluid equations. 

Let us consider now the possibility that the equation for $\mu( \tau)$ tends to a non-zero fixed point. Such a fixed point would be a non trivial solution of NS-Leray, the existence of which is still unsettled, and have been discarded here. 
Note that if a non-zero fixed point exists, it would be a way to find one solution of (\ref{eq:NS-Leray}) by perturbation of solutions of (\ref{eq:Euler-Leray1}). Besides that it would be highly conjectural to say anything more, again because of the lack of known explicit non-trivial solution of  (\ref{eq:NS-Leray}) or (\ref{eq:Euler-Leray1}).  

As two side remarks let us notice first that this breaking of the dilation invariance by the viscosity term could be also operative in the case of direct numerical search of singular solution of the Euler equations because the numerical method always represent imperfectly the original equations. The numerical noise could break the original dilation invariance, that could  interrupt the blow-up by a drift of the dilation parameter, as it happens when a small viscosity is turned on.
Another significant effect of  adding viscosity effects to  the
self-similar solutions  of Euler-Leray equations concerns the dissipation of the energy in the singular domain. Recall that this energy is given by a diverging integral, but if the dilation parameter  $\mu$ tends to zero, the energy (which scales formally as $\mu$) must tend to zero. This paradox could be explained by the spatial spreading of the perturbation $ {\bf{U}}_{c}$ which has length scale increasing as $1/\mu$, 
a situation irrelevant for a
collapse in real flows because large distance coherence should be destroyed
by the field of turbulent fluctuations.

 \section{On the possibility of observing a Leray-like singularity on hot wire records}
 \label{sec:vo}
 The point developed in this paper is that the occurrence of Leray like singularities in flows at high Reynolds number can be put in evidence by measuring the time dependent velocity at a single point. This raises two questions: unless one is very lucky, there is little chance that the point of measurement is exactly the one where the blow-up occurs (this neglecting that there could be no actual blow-up because of the viscosity, a point to which we come back below). Furthermore in the wind-tunnel measurements we shall report on in Sec.~\ref{sec:modane} a mean advection velocity carries along any time dependent event. If the  turbulence intensity is low (turbulent velocity fluctuations small compared to the mean velocity), Taylor's hypothesis of  frozen turbulence can be used to convert temporal experimental measurements into a measurement of a quasi instantaneous space dependence of the velocity. We shall deal with those points now. 

The  self-similar solution of the ``generic type" (\ref{eq:self2})  considered above, which has a dilation invariance, is not the most general solution. 
 Besides the dilation invariance, there are also a translation and a Galilean invariance of solution of the fluid equation, which leads to a more general self-similar solution of the form,
\begin{equation}
{\bf{v}}( {\bf{r}}, t) = (t^*- t)^{-\frac{1}{2}}  {\bf{U}} ( ({\bf{r}} - {\bf{r}}_0 - {\bf{v}}_0 t) (t^*- t)^{-\frac{1}{2}}) + {\bf{v}}_0
 \textrm{,}
 \label{eq:self2.1}
\end{equation}
where ${\bf{r}}_0 $  is the position of the point of measurement, assuming that the singularity occurs at $r=0$ and $t=t^*$.
 This expression represents a flow structure convected with the mean velocity $v_{0}$, which  get an infinite velocity  at $r=r_{0}+v_{0}t^{*}$ at time $t^{*}$.
A local Eulerian probe will record the velocity ${\bf{v}}( {\bf{r}}, t) $ given in Eq.~(\ref{eq:self2.1}) at a given location ${\bf{r}}$ which may be taken as  ${\bf{r}} = 0$. To simplify the expressions let us take also  $t^*= 0$ as time when the singularity (located at $r_{0}$) occurs. 

Consider first cases without advection velocity, namely with $v_0 = 0$. As time goes on, the velocity fluctuation recorded at $r = 0$ can be seen as follows.  As a function of time $t$ the size of the singularity domain decreases, it is of order $(-\Gamma t)^{1/2}$, because $\Gamma\sim u r \sim r^{2}/(-t)$. Therefore when $t$ becomes much smaller than $(r_0^2/\Gamma)$  the singular domain becomes much smaller than $r_{0}$ (its distance to the point of measurement), so that the growth of the velocity close to the singularity cannot  reach the detector.
In other words the growth of the velocity and acceleration will be measured until time $ t \sim -(r_0^2/\Gamma)$. 
As the time delay between $t$ and  the singular time gets smaller than
$(r_0^2/\Gamma)$ the velocity field due to the singularity localized in $r_{0}$  becomes  time independent at large distance $r'\gg r_{0}$, and given by the $\Gamma /\vert r'-r_{0}\vert $ law of  spatial decay,  that gives $\Gamma /r_{0}$ at the detector place. 

When the mean velocity is taken into account the law of decay in space like $1/\vert r'-r_{0} \vert$ becomes a law of decay in time, as recorded by the hot wire, like $\Gamma/\vert r_{0}+v_0 t \vert$. By writing $\vert r_{0}+v_0 t \vert = \vert r_{0}^2 +v_0^2  t^2 + 2 r_0 v_0 t \vert^{1/2}$ and shifting time as  $t = t' - \frac{r_0 }{v_{0}}$ one finds that the signal becomes $\Gamma / \vert v_0 t' \vert $  which is the same for $t' <0$ and $t'>0$. This assumes that at positive (unshifted) time, an Euler-Leray singularity ``bounces" from negative to positive times, its dynamics for positive times (after the singularity) being the same as before the singularity just because of the symmetry of Euler equation under time reversal. However, as shown in section 4 of this paper even a little bit of viscosity should yield a strong asymmetry of the time signal because it makes vanish the singularity in the scale of the logarithmic time.

The standard view on measurements of velocity fluctuations in wind tunnels by hot wires is that the only thing one can observe is the space dependent part of the velocity field because the typical time of evolution of the turbulent fluctuations is much longer than the typical time of advection of the structure, just because those two times are related to velocities by the simple formula $r/v$ so that the bigger velocity yields the shortest time for a given distance $r$. This is called Taylor assumption/hypothesis  of frozen turbulence. In the Modane experiments described in Sec.~\ref{sec:modane}, the standard deviation of $v$ is  smaller than $v_{0}/10$,   but the maximum amplitude  of the velocity fluctuations ($v_{max}-v_{min}$) is of the same order as or even bigger than the mean advection velocity $v_{0}$. Therefore applying Taylor assumption is not that obviously permitted. 
Let  us return to the scaling relation (\ref{eq:cubvb}) and precise the role of the advection in the time derivative the velocity, an important point  to compare the scaling laws with experimental data (see next section).  From (\ref{eq:self2.1})  we get
 \begin{equation}
\frac{d{\bf{v}}( {\bf{r}}, t)}{dt} =\frac{1}{ (-t)^{3/2}} \left(U + R \frac{\partial U}{\partial R}\right) + \frac{v_0}{(-t)} \frac{\partial U}{\partial R}
{.} 
 \label{eq:advection}
\end{equation}
 On the r.h.s. of (\ref{eq:advection})  the leading order term is the first one (as $t$ tends to zero). The second  term, coming from the advection, is  like $1/(-t)$. To be more precise,  let us compare the order of magnitude of these two contributions to the the acceleration. The first one is of order $\Gamma^{1/2} /(- t)^{3/2}$ whereas the second one  which was used to derive (\ref{eq:cubvb}), is of order $v_0/(- t)$. When $v_0$ gets very big the advection effect can dominate over the self-similar dynamics but it is not always dominant, since
the first term leading to the scaling law (\ref{eq:cubv}) becomes dominant as $t$ gets closer and closer to $0$.

We show below that  the experimental observations agree well with the relationship (\ref{eq:cubv}) between velocity and acceleration where the advection  effect on the measurements is neglected, although the experiment do not fit  the relation (\ref{eq:cubvb})  deduced with the Taylor hypothesis. In summary, the  assumption of  Taylor frozen turbulence has to be used with caution when looking at extreme events.

  \section{Analysis of wind tunnel records.}
 \label{sec:modane}
 We tested the two  relations (\ref{eq:cubv})-(\ref{eq:Kolm})  and also (\ref{eq:cubvb})-(\ref{eq:Kolmb})  against experimental results by comparing the values of the velocity fluctuations $u=v-v_{0}$, and of the acceleration recorded at the same place and same time,  $\gamma_i= (v_{i+1}-v_{i}) f$ , with $f$  the sampling frequency.
 We looked  at the data obtained in the S1MA wind-tunnel from ONERA in Modane, where the turbulent velocity was recorded by  hot wires. The first  subsection below is relative to data taken in the return vein  of the tunnel   in the $90$s \cite{expmod}-\cite{exp2mod}.  Subsection \ref{sec:modane2} uses recent  measurements  made in 2014 in the framework of a ESWIRP European project, also  in the  wind-tunnel of Modane \cite{mickael}. 

 Our aim was to use experimental data in order to conclude about the presence of self-similar solutions  in the turbulent flow, and more precisely if self-similar solutions of type (\ref{eq:self2}) do show-up. If  they do, even as rare events, they should be seeable at least for large $\gamma$ and  $u$ values, where one expects a relation  of type (\ref{eq:cubv}), or (\ref{eq:sedov}), between acceleration and velocity fluctuations. On the contrary, if  Kolmogorov-scaling rules  the dynamics, large accelerations  (resp. velocity) should occur when the velocity fluctuations (resp. acceleration) are small.  
 
 \subsection{ Data taken in the return vein in the ${\bf{90}}$s. } 
 \label{sec:modane1}
We first present our study of  a $10$ min. record  of the wind velocity, taken at  sampling frequency $f=25$Khz   ($\mathcal{N}=13.7 $ millions of points in time) by a single hot wire  located in the return vein of the tunnel. The mean wind  velocity was $20.55$m/s, with standard deviation $1.7$m/s.  The Reynolds number  $Re_{\lambda}=\sqrt{15Re}$ is about 2500 (one of the largest value in this kind of experiment). The  ``acceleration''  in these data (as defined above) have standard deviation $\sigma_{\gamma}=1803m/s^{2}  $, and the maximum acceleration was about $30$ times this value, reaching about $5000$ times the gravitational constant.
In order to test if the scalings associated to self-similar solutions can be extracted from the experimental data, we have studied the behavior of  two set of conditional moments, the first one given $\gamma$, which is presented just below, the second one given the velocity $u$, see subsection \ref{sec:cond-v}.

 \subsubsection{Moments conditioned on acceleration}
 \label{sec:cond-accel}
  From the set of paired  values $(v_{i}, \gamma_{i})$, $i=1..\mathcal{N}$,  taken at the same time and same place, we first look at conditional moments  of the velocity fluctuations given the acceleration, noted  $\left<u^{n}\right>_{\gamma}$, with $n=1,2,3$.
\footnote{We use the notation $\left<u^{n}\right>_{\gamma}$   for  a conditional moment  given $\gamma$,  instead of the standard notation  $ \left<u^{n}\vert {\gamma} \right>$  to avoid confusion with the ratio $\left<u^{n}/\gamma \right>$ also used below.}  Let us compare the scalings for  Euler-Leray and Kolmogorov predictions in terms of the size of the structures. For Leray case self-similar solutions close to $t^{*}$ are obviously associated to short lived  and small-sized spatial structures, as seen from (\ref{eq:self})  which gives, 
 \begin{equation}
\left<u^{3}\right>_{\gamma} \sim \Gamma \gamma \sim 1/r^{3}  \;\; \; \textrm{or} \;\;\;  \left<u^{2}\right>_{\gamma} \sim\frac{ \Gamma }{v_{0}} \gamma_{\rm Taylor} \sim 1/r^{2}  \qquad (Leray)
\textrm{.} 
 \label{eq:moment3}
\end{equation}
The two possibilities in (\ref{eq:moment3}) correspond to the relations (\ref{eq:cubv}) and (\ref{eq:cubvb}), the former is deduced with $\gamma= \partial u/\partial t$, the latter is based on Taylor hypothesis, $\gamma \sim v_{0}\partial u/\partial x$. In addition, we note that for singularities associated to self-similar solutions with Sedov-Taylor exponents, the above relations becomes  $\left<u^{8/3}\right>_{\gamma} \sim 1/r^{4} $.
Such singular events associated to small $r$ values, may be identified by  large  values of acceleration \textit{and} velocity fluctuation. On the other hand  Kolmogorov scalings (\ref{eq:Kolm})-(\ref{eq:Kolmb})  predict that small scales  are also connected to large acceleration but they are linked to small velocity fluctuation according to the rule,
  \begin{equation}
\left<u\right>_{\gamma} \sim r^{1/3}  \; \; ; \; \  \gamma \sim r^{-1/3}  \;\;   \textrm{or}   \;\; \gamma_{\rm Taylor } \sim r^{-2/3}  \qquad (K)
\textrm{.} 
 \label{eq:moment4}
\end{equation}
As above the two possibilities depend on the validity of Taylor hypothesis, they correspond  to (\ref{eq:Kolm}) or (\ref{eq:Kolmb}) respectively .
 
If Leray-like solutions are formed in the flow, they have to co-exist with Kolmogorov fluctuations, then
 large acceleration events  can appear either with a large velocity ( due to Leray condition),  or with a small velocity ( due to Kolmogorov scalings). 
 
 Therefore if one observes a linear relation between $\gamma$ and $\left<u^{3}\right>_{\gamma}$ (or  $\left<u^{8/3}\right>_{\gamma}$) one could expect that the ratio $\left<u^{3}\right>_{\gamma}/\gamma$ (or  $\left<u^{8/3}\right>_{\gamma}/\gamma$)would  be smaller than the true value of  the circulation $\Gamma$ (or energy $E$)  around the singular point, because this ratio should result from  a kind of competition between the two processes of building small scale fluctuations.

  The conditional moments  deduced from the experimental data are defined formally as
\begin{equation}
\left< u^{n}\right>_{\gamma}=  \int{ u^{n}     P_{\gamma}(u) du     }
\textrm{,} 
 \label{eq:moment}
\end{equation}
where $P_{\gamma}(u)$ is the  the conditional probability  of the velocity  fluctuation for a given value $\gamma$ of the acceleration which is deduced from the join probability $P(u_{i},\gamma_{j})du d\gamma $  
  for the pair of variables $(u,\gamma)$ to be inside the domain $ (u_{i},u_{i}+du)\times (\gamma_{j}, \gamma_{j } + d\gamma)$. From the raw data it is given by the number of points recorded in this domain divided by the total number of recorded points, 
 \begin{equation}
P(u_{i},\gamma_{j}) du d\gamma=  N_{i,j} /\mathcal{N}
\textrm{.} 
 \label{eq:pij}
\end{equation}

The conditional probability $ P_{\gamma_{j}}(u_{i}) du$ for the velocity to be inside the interval $[u_{i},u_{i}+du]$ given the event that acceleration  is inside $ [\gamma_{j}, \gamma_{j } + d\gamma] $, is  given by   $ N_{i,j} / N_{j}$, where $N_{j}=\sum_{i} N_{i,j}$. Using (\ref{eq:moment})-(\ref{eq:pij}), we get  the  following expression for $\left< u^{n}\right>_{\gamma}$ in terms of the number of points recorded in the elementary domains, 
\begin{equation}
\left< u^{n}\right>_{\gamma}=  \sum_{i} u_{i}^{n} \frac {N_{i,j}} {N_{j}}
\textrm{.} 
 \label{eq:moment2}
\end{equation}

\begin{figure}
  \centerline{ (a)\includegraphics[height=1.75in]{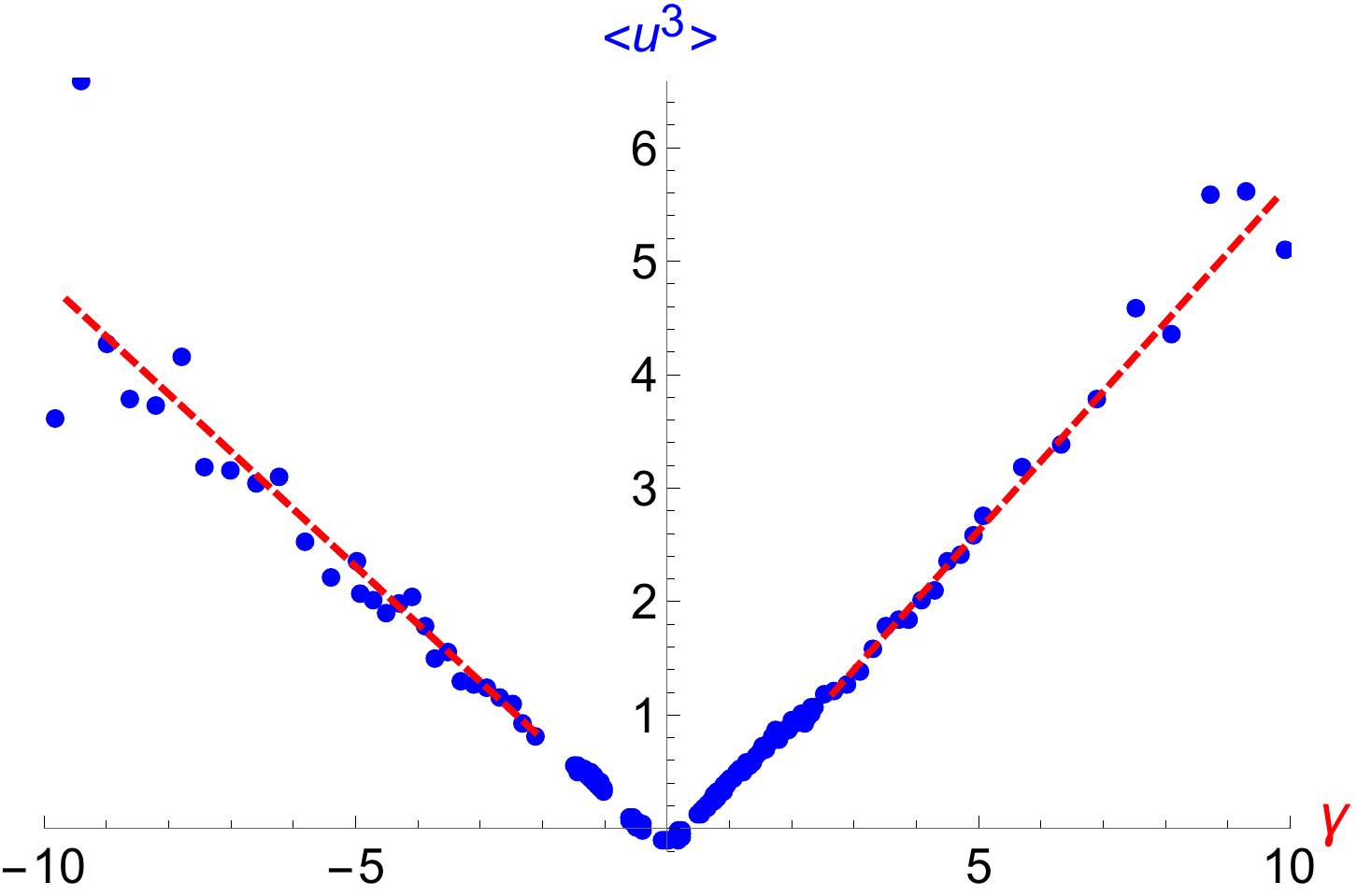}
(b)\includegraphics[height=1.75in]{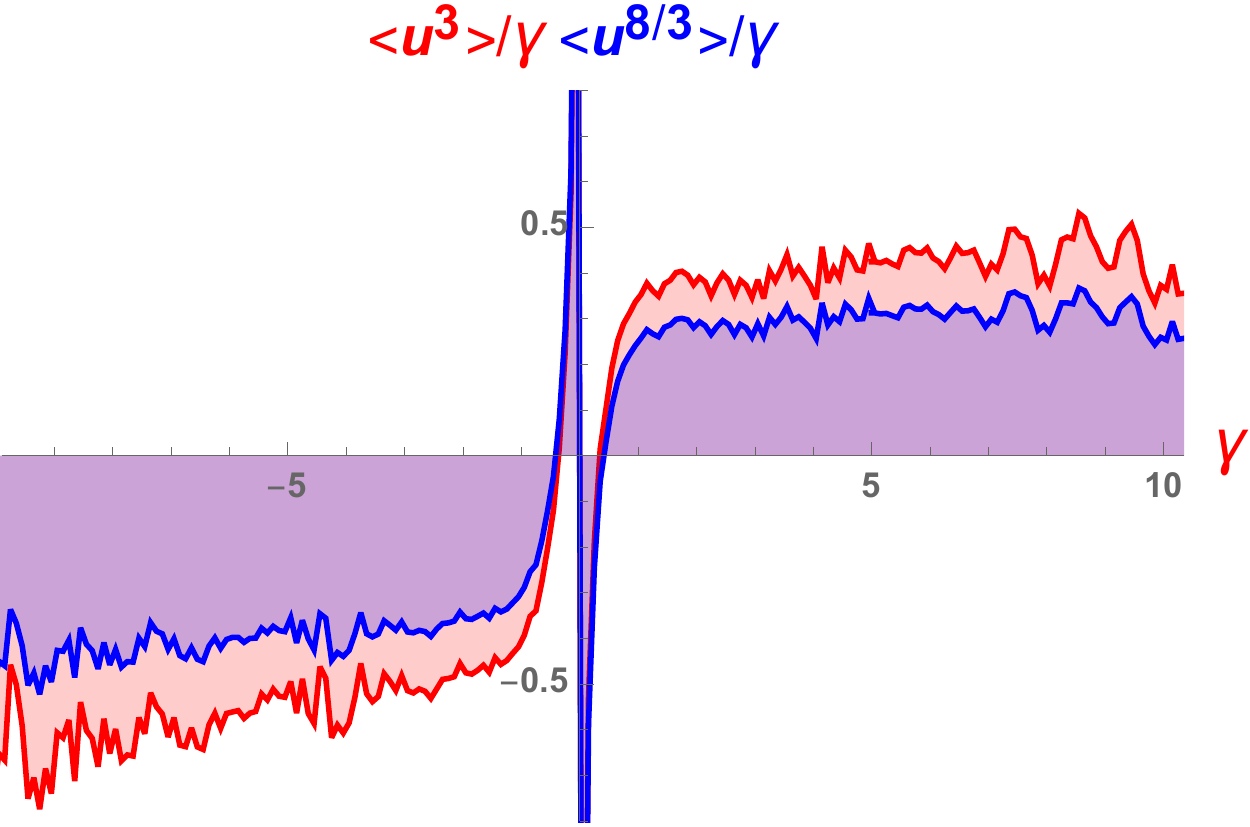}}
\caption{  Experimental  test to investigate the validity of  the scalings associated to self-similar solutions of the Euler equations.  In (a) the conditional average  $\left<u^{3}\right>_{\gamma}$  (for a given value of $\gamma$) versus $\gamma$ agrees with (\ref{eq:cubv}).  Curves (b) (drawn with filling to axis)  compare the  data with predictions  of  relations (\ref{eq:cubv})  and (\ref{eq:sedov}), red and blue curve respectively. The ratio $\left<u^{3}\right>_{\gamma}/\gamma$   in red increases slightly with $\gamma$ with  a quasi-plateau for $\vert \gamma\vert \gtrsim 1.5 \sigma_{\gamma}$, the physical value of the constant $ \Gamma$ is given in the text.  The blue curve shows that the ratio  $\left<u^{8/3}\right>_{\gamma}/\gamma$  which displays a cleaner plateau agrees well with  the Sedov-Taylor relation (\ref{eq:sedov}). In the figures $u$ and  $\gamma$ are  in units of their  respective standard deviation or rms. The bin width for the velocity is $\delta u=0.5$, for the acceleration $\delta \gamma$  increases  from the origin to the edges ( from $0.01$ to  $0.5$) as indicated by the interval between the points in (a).}
\label{fig:Leray}
\end{figure}

\begin{figure}
  \centerline{ 
(a)\includegraphics[height=1.75in]{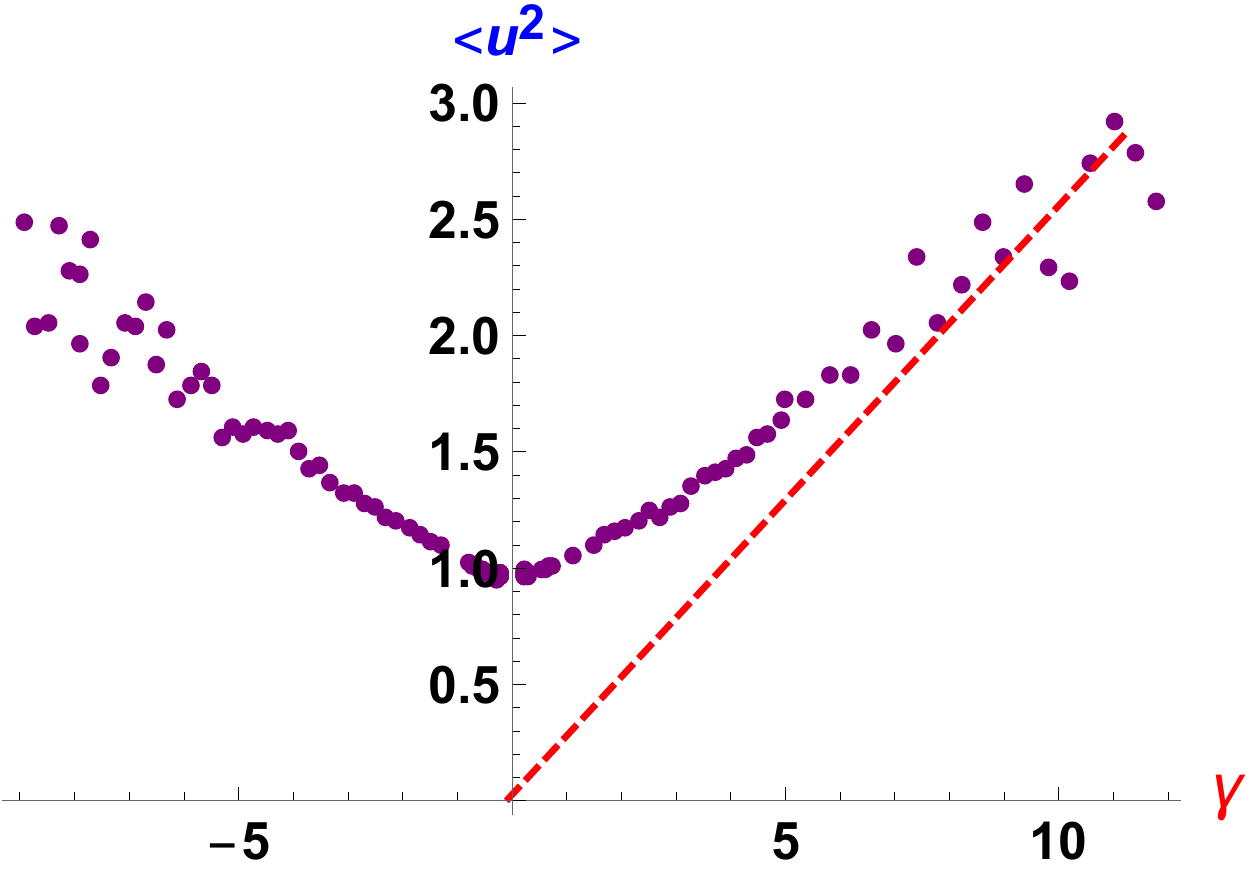}
(b)\includegraphics[height=1.75in]{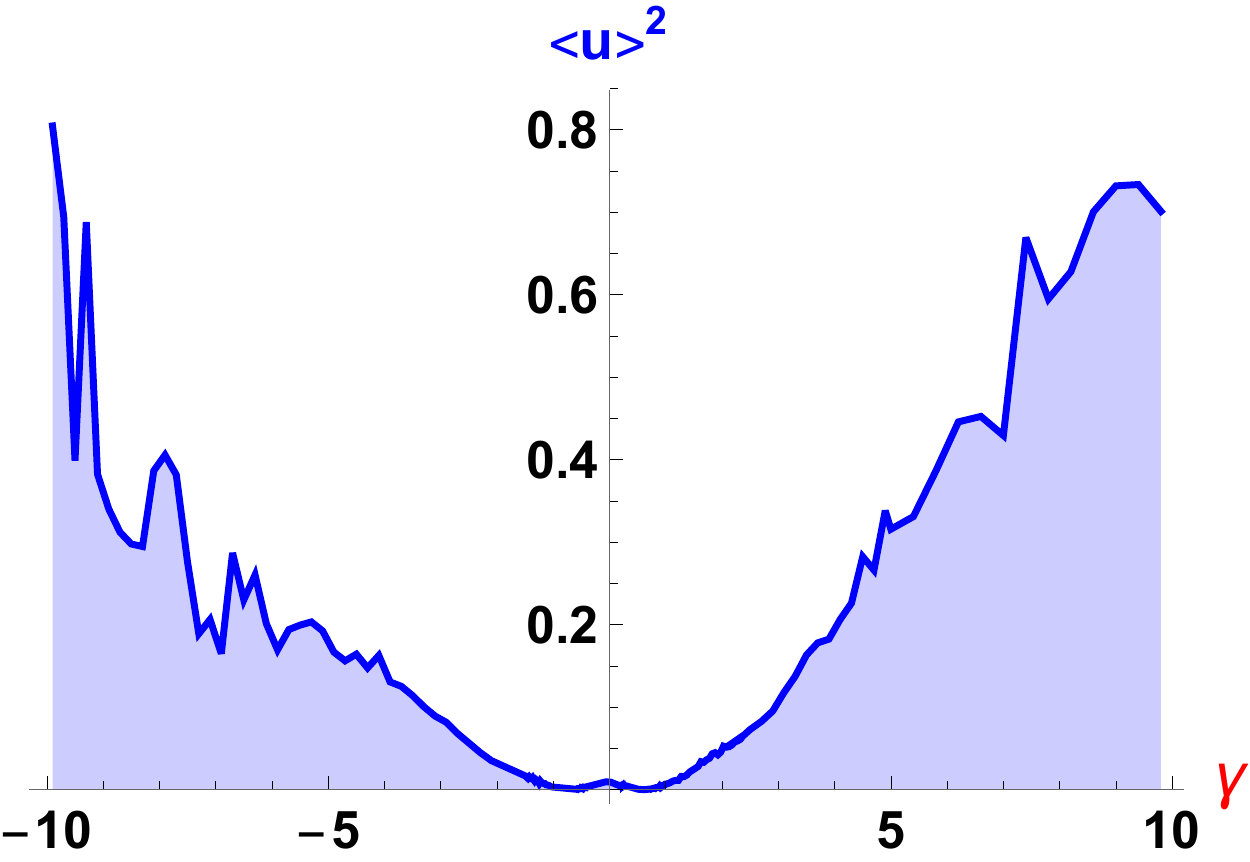}
  } 
  
\caption{   Experimental  test which shows that the scaling (\ref{eq:cubvb}) derived from Euler-Leray solutions within Taylor hypothesis is not good. (a) displays the conditional moment $\left<u^{2}\right>_{\gamma}$  versus $\gamma$. The dashed line indicates a short range where eventually a linear relation exists between  $ \left<u^{2}\right>_{\gamma}$ and $\gamma$, that corresponds to less than $0.5$ point per thousand,  see  the insert of Fig.(\ref{fig:kolm})-(b).  Curve  (b) filled to axis, displays  $ (\left<u\right>_{\gamma})^{2}$ versus $\gamma$ which is also non linear.  In the figures $u$ and  $\gamma$ are  in units of their  respective standard deviation. Same bin widths as in previous figures.}
\label{fig:Leray2}
\end{figure}

\begin{figure}
  \centerline{ 
(a)\includegraphics[height=1.5in]{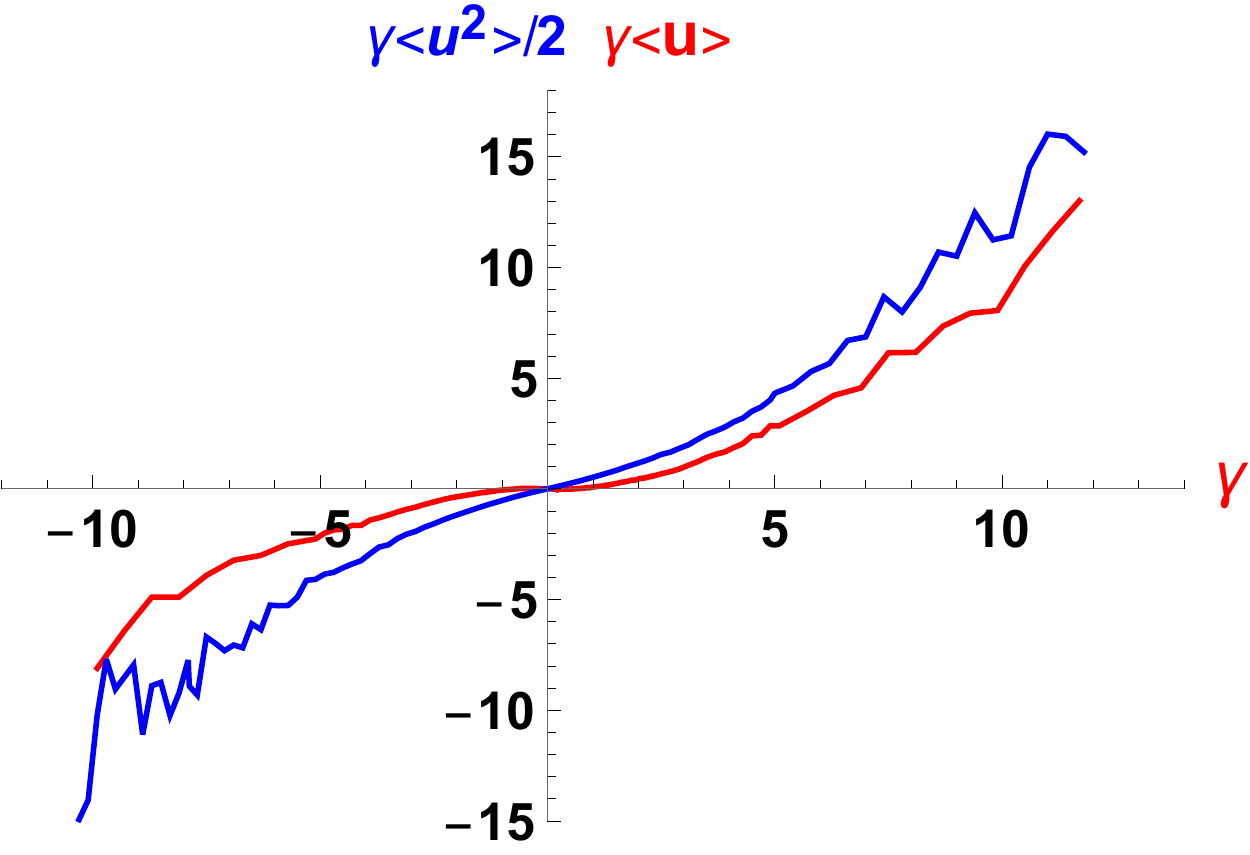}
(b)\includegraphics[height=1.5in]{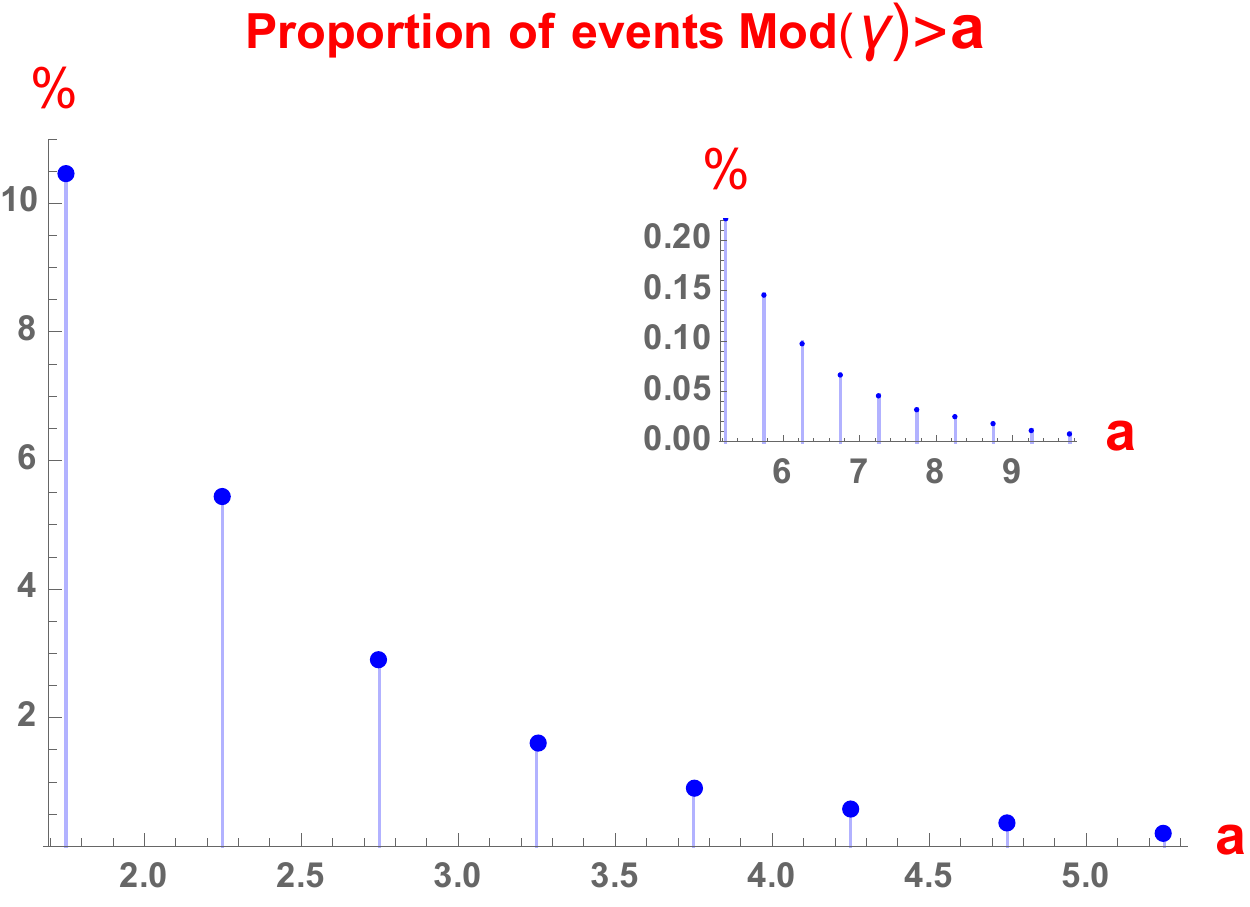}
  }
\caption{(a)  Experimental  test  for Kolmogorov scalings (\ref{eq:Kolm})  and (\ref{eq:Kolmb}).  Conditional moments  $ \gamma\, \left<u\right>_{\gamma}$ (red curve) and  $1/2 \gamma\left<u^{2}\right>_{\gamma}$ (blue curve) given  $\gamma$.  The factor $1/2$ in front of  $\gamma\left<u^{2}\right>_{\gamma}$  is set to make  easier the comparison between the behavior of the  two curves. (b) Percentage of rare events with acceleration larger than $a$ (in units of standard deviation). Same bin widths as in previous figures. }
\label{fig:kolm}
\end{figure}

To see which one, if any, of the relations  (\ref{eq:cubv})-(\ref{eq:sedov})  or (\ref{eq:cubvb}) one hand hand, and (\ref{eq:Kolm})  or (\ref{eq:Kolmb})  on the other hand, agrees with the experimental data, we plot  in Figs.(\ref{fig:Leray}) and  (\ref{fig:Leray2}) respectively  the observed values of  $\left<u^{3}\right>_{\gamma}$,  and $\left<u^{2}\right>_{\gamma}$ and we show in Fig.(\ref{fig:kolm})-(a) that Kolmogorov relations do not fit the data. 

  Fig.\ref{fig:Leray}-(a) shows that  in average, the power $3$ of the velocity fluctuations increases quasi-linearly with the acceleration, in agreement with the  relation (\ref{eq:cubv}). This statement  is completed by Fig. \ref{fig:Leray}-(b)  which displays the ratio $\left<u^{3}\right>_{\gamma}/\gamma$, red curve. Because this ratio  slightly increases with $\gamma$ at  large values of $\gamma$,  we plot  on the same curve the ratio corresponding to Sedov-Taylor scalings,  $\left<u^{8/3}\right>_{\gamma}/\gamma$.  Although the exponents $3$ and $8/3$  are  very close, we have to remark that  the blue curve displays a  clearer flat behavior than the red one. In both cases the  constant (or quasi-constant) behavior extends on a wide range of order $\vert \gamma \vert \gtrsim 1.5\sigma_{\gamma}$.
Differently Figs. \ref{fig:Leray2} show that Leray relation (\ref{eq:cubvb}) with Taylor hypothesis  does not fit so well the data, because the domain where $u^{2}\propto \gamma$ is very short or non-existent, see captions.
The latter  poor fit illustrates that the  events associated to  large acceleration and  large  velocity fluctuations  are beyond the validity of Taylor hypothesis.

In summary the  experimental data agree well with our hypothesis of existence of Leray-type singular events  in turbulent flow.  We observed a good enough fit between the Leray's scalings (\ref{eq:cubv})  and the  experimental conditional moment, which behaves as $\left<u^{3}\right>_{\gamma}\sim \gamma$. Surprisingly we note an even better fit  when comparing the data with the  relation (\ref{eq:sedov}) associated to Sedov-Taylor exponents, as illustrated in  Fig.(\ref{fig:Leray})-(b). The  large domain spanned by these  promising fits  is a striking  result which is even a bit unexpected.  It implies that the prefactor $\Gamma$ in (\ref{eq:cubv}) or $E^{1/3}$ in (\ref{eq:sedov}),  is not changing much from one singular event to the other, and that eddies of different size do not change this relation on average.  

In the following we compare  Leray's  scalings with the experiment in order to support our theory, making the hypothesis that the circulation or the energy is conserved in the singular domain.

Fig.(\ref{fig:kolm})-(a)  shows  that the Kolmogorov scalings leading to relations (\ref{eq:Kolm})-(\ref{eq:Kolmb}) do not agree  with the data because  no plateau shows-up in the dependence of the products $ \gamma  \left<u\right>_{\gamma} $  and $ \gamma  \left<u^{2}\right>_{\gamma} $ with respect to $\gamma$. In  particular those quantities strongly increase for large  accelerations. This result is in favor of the occurrence of Leray singularities in the flow. Furthermore it shows that the singular structures have a stronger effect on the moments than other kind of fluctuations (called ``normal" later on).  In the large acceleration domain the effect of  normal fluctuations,  would be to lower the 
 circulation or energy value around singular points, since  small eddies  are associated 
to small $u$ values for Kolmogorov scalings, as written in (\ref{eq:moment4}).  
Therefore one expects that the  value of the slope in Fig.\ref{fig:Leray}-(a)  is  smaller than the real value of the circulation close to a singular point. In physical variables the circulation is $\Gamma= s \sigma_{v}^{3}/ \sigma_{\gamma}$ where $s$ is the slope of  curve (a), or the height of the plateau in (b).
From the observed value  $\Gamma_{\gamma}=1.6$ $10^{-3}$ $m^2/s$ of the circulation, one may find the local Reynolds number  $Re_{\gamma}=\Gamma_{\gamma}/\nu$, which is about $160$ (taking the kinematic viscosity of air about $\nu \simeq 10^{-5}$ $m^2/s$ at room temperature). It is a large but not very large  Reynolds number, see the discussion in next subsection. 

Finally let us emphasize  that  there is about $5$  to $10$ per cent of  points in the  whole record  which agree with the Leray's scaling  (\ref{eq:cubv}) or (\ref{eq:sedov}), namely which correspond to acceleration  values (scaled to $\sigma_{\gamma}$) in the domain $\vert \gamma \vert  \gtrsim 2 $, see Fig.(\ref{fig:kolm})-(b),  where Leray's  or Sedov's scaling  is observed. If such events are really associated to singular solutions, this should indicate that the formation of   self-similar  solutions of the  Leray-type is not so rare.

\subsubsection{ Moments conditioned on velocity.}
\label{sec:cond-v}
\begin{figure}
\centerline{ 
(a)\includegraphics[height=1.5in]{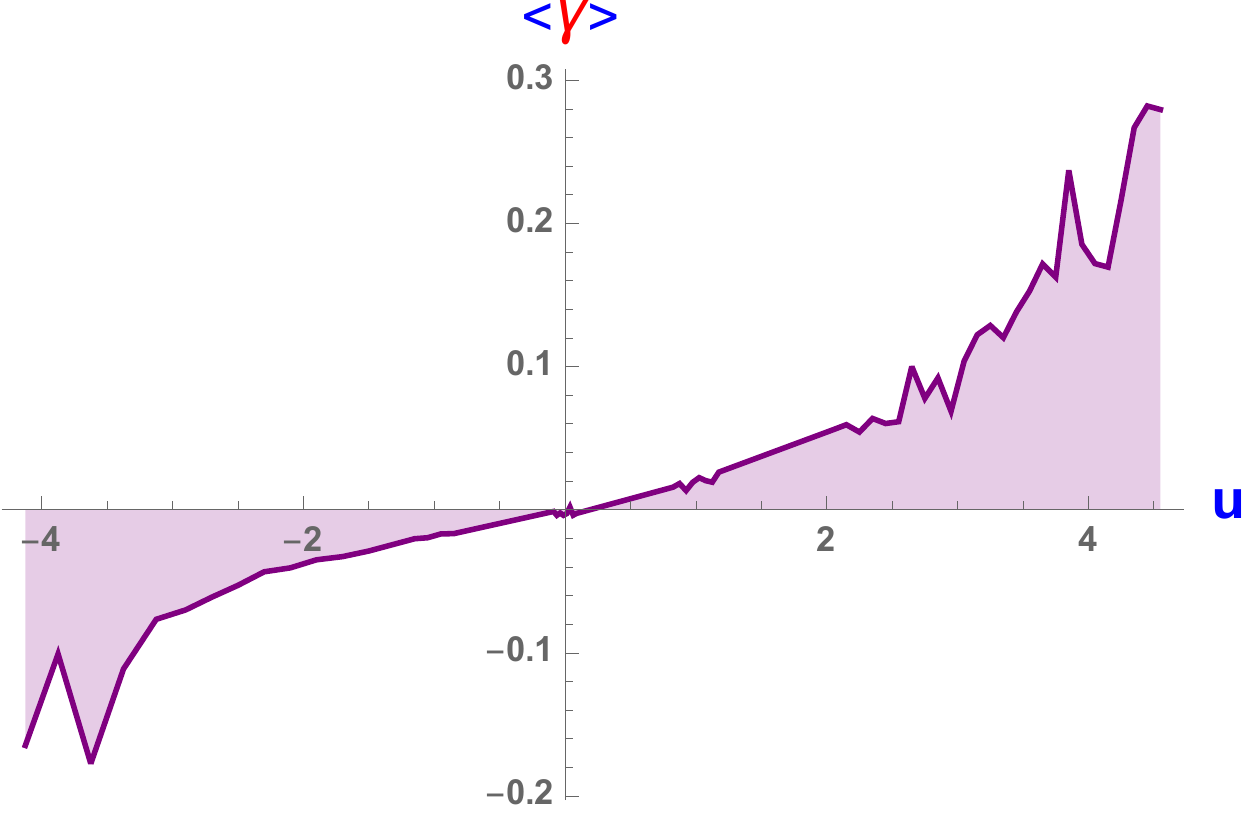}
(b)\includegraphics[height=1.5in]{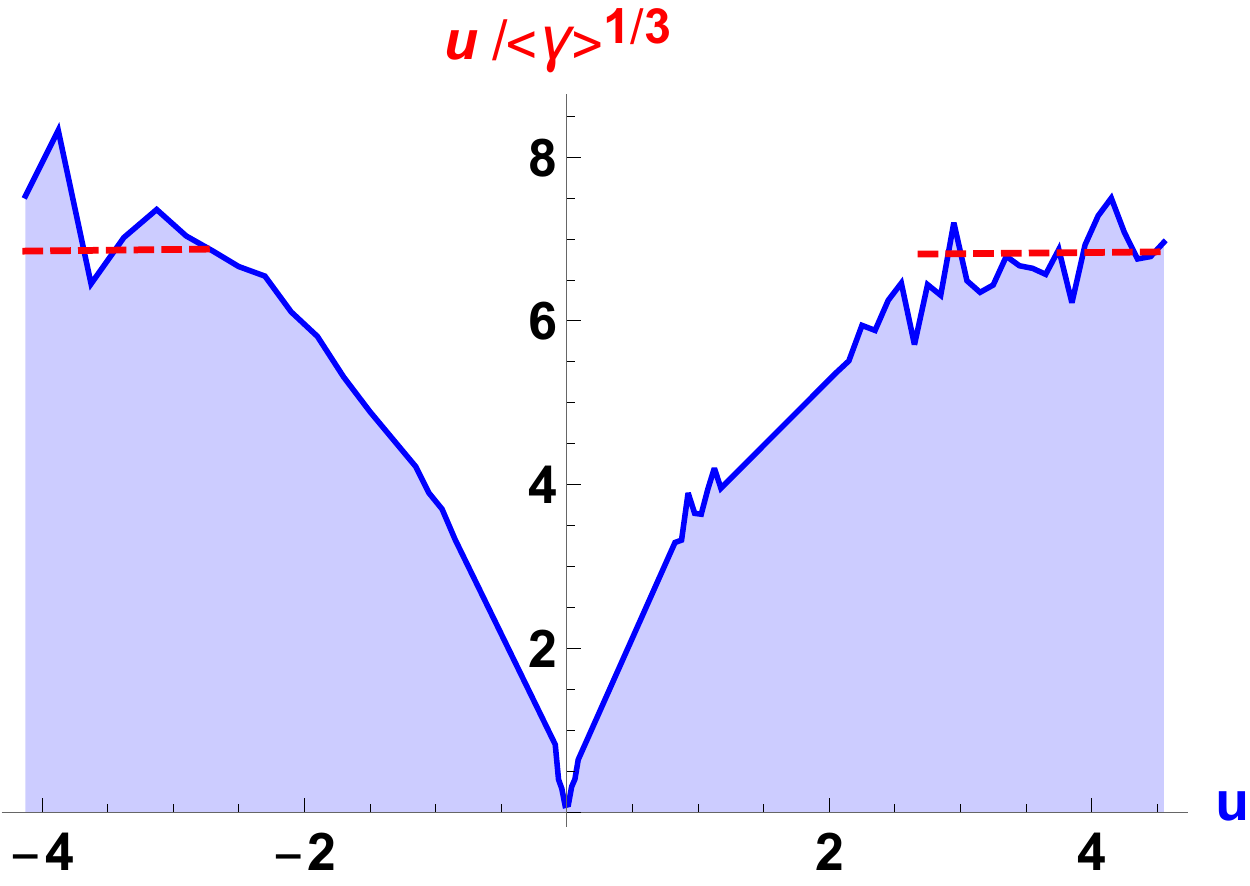}
  }
\caption{ (a)Experimental plot of  the conditional moment given $u$, $ \left<\gamma\right>_{u} $,  versus $u$, both  in units of their rms.  In (b)  the ratio  $ u/\left<\gamma\right>_{u}^{1/3} $  versus $u$  displays a  plateau for  $u \gtrsim 2.5 \sigma_{u}$.
}
\label{fig:condv}
\end{figure}

Symmetrically we  have also investigated the behavior of  $\left<\gamma\right>_{u}$,  the conditional moment (average value of $\gamma$)  given a velocity fluctuation $u$. A large velocity $u$ is expected  for small spatial scales with Leray's scaling (\ref{eq:moment3}), and for large scales with Kolmogorov scalings (\ref{eq:moment4}). Therefore peaks of $u$ can be due  either to  the presence of singular events of small size,  if they exist, or associated to large ``normal'' structures (K relation). But for such large eddies, large velocities should occur with small acceleration, contrary to singular events which are  related to large acceleration (as explained  in section \ref{sec:cond-accel}). The result of this study is 
shown in Figs. \ref{fig:condv}  (curves drawn with filling to axis).

 In Fig.\ref{fig:condv}-(a) the curve shows that large values of $\vert u\vert$ are associated to maxima of $ \left<\gamma\right>_{u}$. This result is in agreement with Leray's predictions and in contradiction with Kolmogorov's ones. But we have to notice that the values of $ \left<\gamma\right>_{u}$ fitting Leray's relation are small (in physical units they are smaller than the standard deviation $\sigma_{\gamma}$).  This is explainable by the competition between singular events and  large ''normal'' eddies, both contributing to large velocities, as resumed in equations (\ref{eq:moment3})-(\ref{eq:moment4}).  
 This drastic reduction of the observed $ \left<\gamma\right>_{u}$  with respect to what is expected if singular events are not in competition with large eddies, points out that ''normal'' fluctuations contribute noticeably to what happens  at large velocity. To precise in what domain Leray's scalings win Kolmogorov's ones, we show in Fig.\ref{fig:condv}-(b)  the ratio  $ u/\left<\gamma\right>_{u}^{1/3} $, which is approximately constant  for velocity $u$ larger than about $ 2.5 \sigma_{u}$.  In this domain the data are in agreement with Leray's scalings (\ref{eq:cubv}), however we note that the plateau is narrower than the one of Fig.\ref{fig:Leray}-(b) for the previous study of  conditional momenta given  acceleration.  Comparing with the previous subsection, this result could yield that  the contribution of large  normal eddies is more active  to reduce $\left<\gamma\right>_{u} $ , than the contribution of small normal eddies to reduce the moment  $\left<u^{3}\right>_{\gamma} $  calculated in the previous subsection. 
 
 Because of the small value of $ \left<\gamma\right>_{u}$, the apparent circulation $\Gamma_{u}$ and the local Reynolds number $Re_{u}$  are greatly enhanced with respect to the  values of the corresponding quantities in Sec.~\ref{sec:cond-accel}. Here  we get  $Re_{u}\sim 10^{5}$, which is  of order of  the Reynolds number in  Modane experiment (where $Re = 4.2\; 10^{5}$).
Due to the huge discrepancy between  the Reynolds number deduced by the two methods described in  Sec.~\ref{sec:cond-accel} and Sec.~\ref{sec:cond-v}, the local Reynolds number at the singularity cannot be fairly estimated, nevertheless we can assert that it is much larger than unity which is the typical value around small eddies in the dissipative range.

 Note that we have used the Eulerian definition of the acceleration, $\gamma_{E}= \frac{\partial{\bf{u}}}{\partial t}$,  to compare  Leray  and  Kolmogorov scaling laws with  the experiments. This is correct for Leray scalings because the self-similar solution is derived with the hypothesis that the two terms of the Lagrangian acceleration $\gamma_{L}= \frac{\partial{\bf{u}}}{\partial t}+ {\bf{u}}\cdot \nabla {\bf{u}}$, are of same order. Actually  Kolmogorov scalings laws should be written as    $u\gamma_{L} \sim \epsilon$,  or $u^{2}\gamma_{L} \sim  v_{0}\epsilon$ (if Taylor's hypothesis is assumed). 
Because the experimental data  display a strong increase of  $\left<u^{2}\right>_{\gamma_{E}}$ and $\left<u\right>_{\gamma_{E}}$   as $\gamma_{E}$ increases,  it is very unlikely that
 both quantities become functions decreasing  like $1/\gamma$, if Lagrangian acceleration data were used  in place of the Eulerian acceleration used here.
We conjecture  that  scaling laws could not  make so large difference  between two quantities which represent the same thing.

In summary the  record  taken in the return vein of the wind tunnel agrees with the predictions of our analysis, based on the existence of self-similar solutions in the turbulent flow.  We observed that  in average the events with large acceleration  are  associated to large velocities fluctuations. The statistical study using  probability  distributions conditionally to a given value of acceleration, and of velocity fluctuation, shows  that on average, there is a linear relation between $\gamma$ and  $u^{z}$ with $z\simeq 3$, for large $\gamma$ and $u$. This proves that  singular events (if they exist) are not rubbed out by small eddies contributions in the former case (Sec.~\ref{sec:cond-accel}),  and  by large structures in the latter (this subsection). On the contrary singularities show up as 
winning in the competition with  normal eddies  (small and large ones), since the linear relation  between $u^{z}$ and $\gamma$ is verified experimentally  in averaged, for a large range of acceleration  and velocity values.

 \subsection{2014-data : grid turbulence} 
 \label{sec:modane2}
\begin{figure}
\centerline{
(a) \includegraphics[width=2 in]{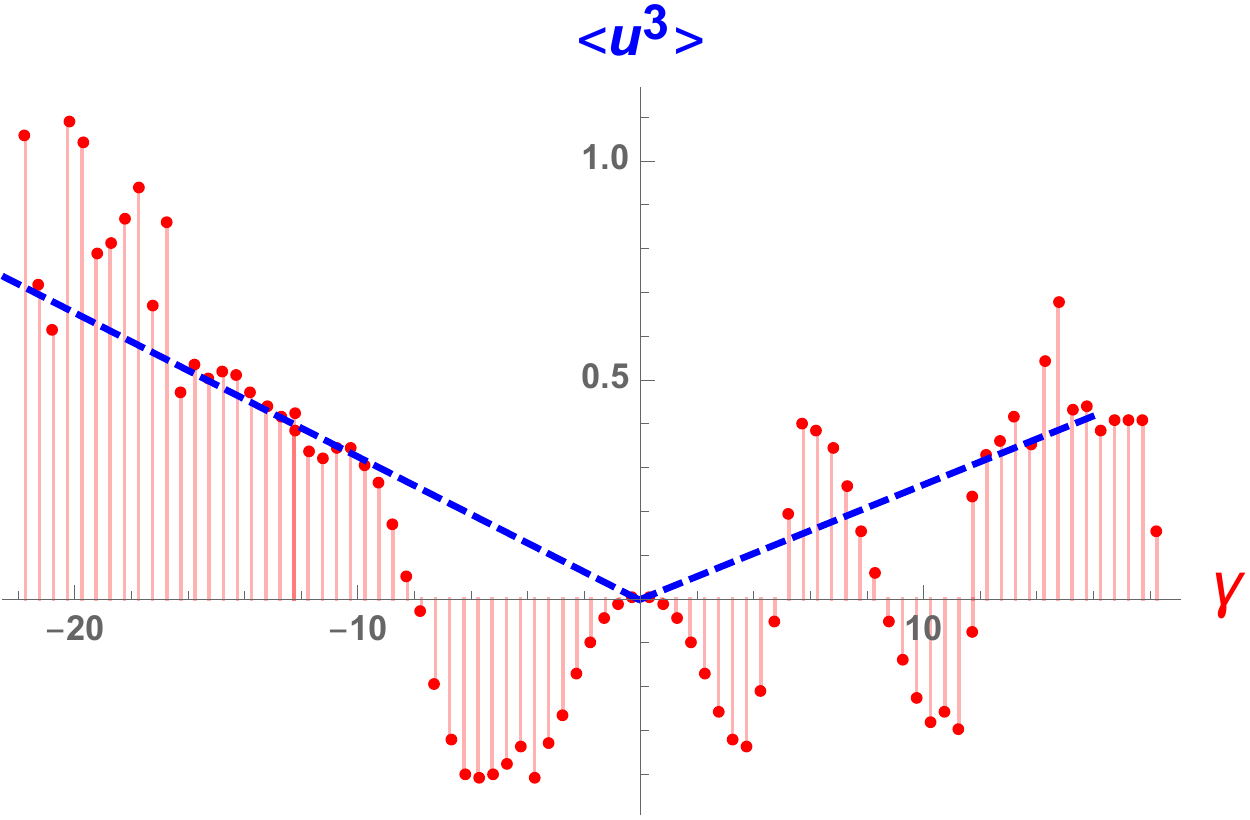}
(b) \includegraphics[width=2 in]{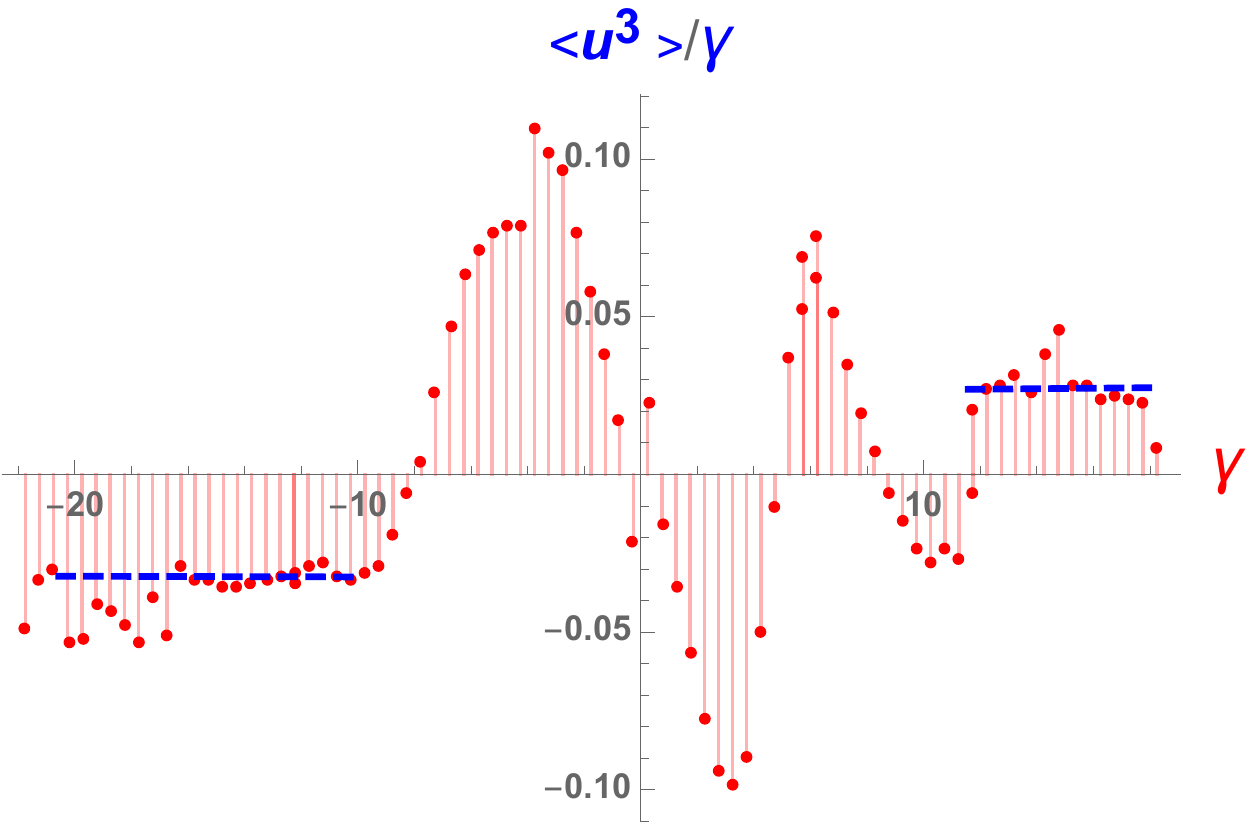}
}
 \caption{ Conditional moment  calculated from data recorded far behind the grid , (a)$\left<u^{3}\right>_{\gamma} $, (b) $\left<u^{3}\right>_{\gamma} /\gamma$  versus $\gamma$, in units of their rms. The mean velocity  and standard deviations are $\left<v_{0}\right> = 40.6$m/s,   $\sigma_{v}=0.9$m/s and $\sigma_{\gamma}=11000$m/$s^{2}$.  The bin widths are $\delta u=\delta \gamma =0.5$ in units of their rms}
 \label{fig:23m}
\end{figure}

\begin{figure}
\centerline{
 \includegraphics[width=2in]{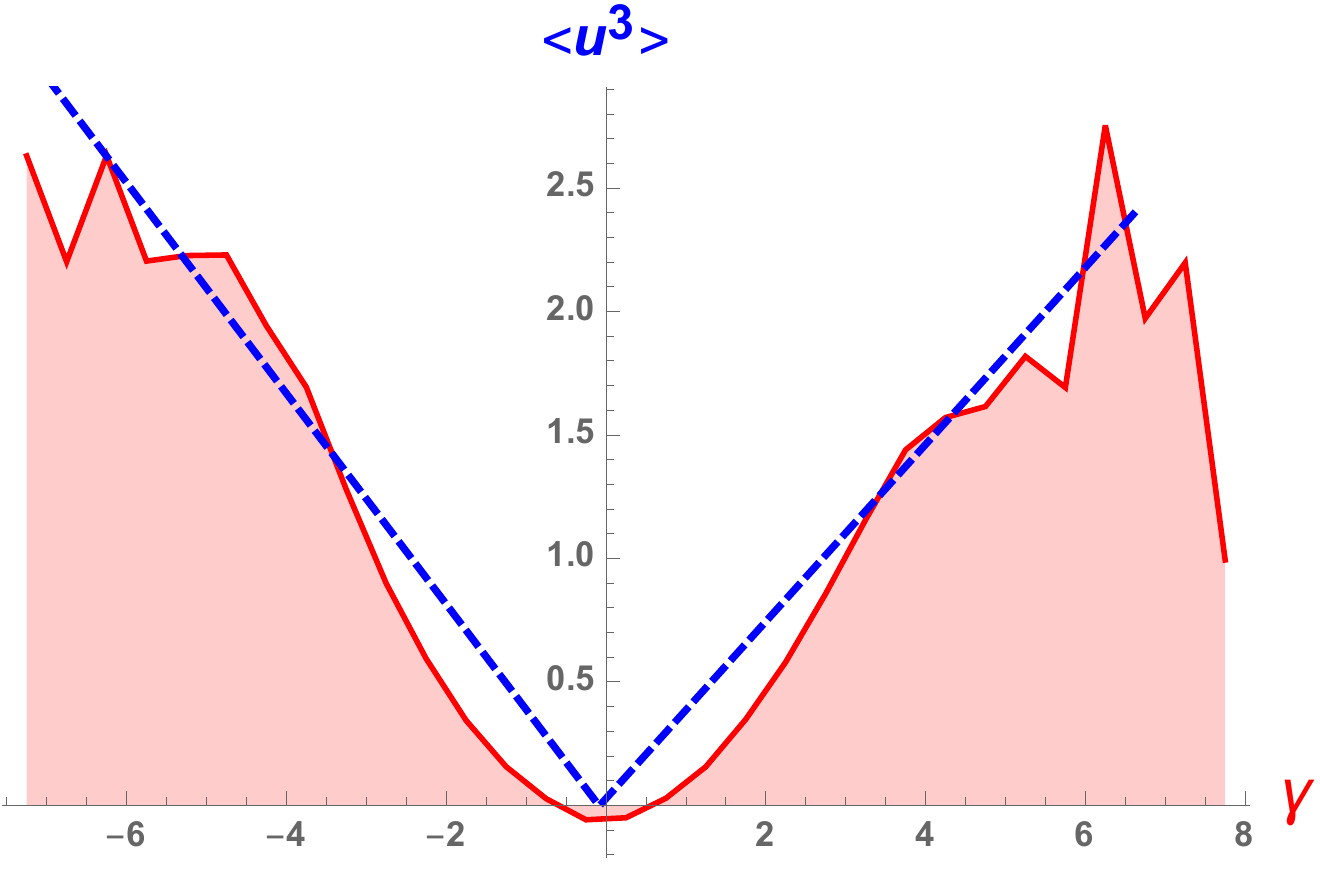}
  \includegraphics[width=1.8in]{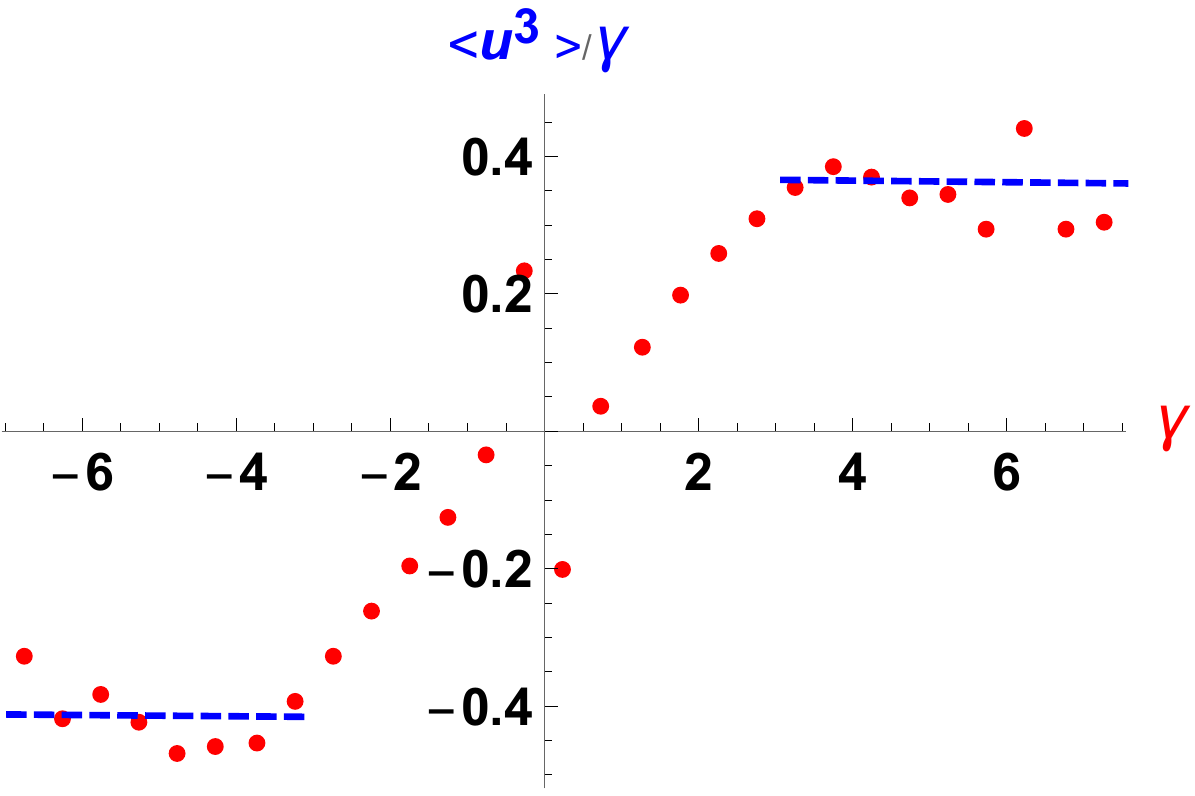}
   }
    \caption{ Conditional moment for data recorded  at $8m$ behind the grid, (a) $\left< u^{3} \right>_{\gamma} $, (b) $\left<u^{3}\right>_{\gamma} /\gamma$  versus $\gamma$, in units of their standard deviations. The mean velocity  and standard deviations are $ v_0 =43$m/s,   $\sigma_{v}=2$m/s and $\sigma_{\gamma}=38250$m/$s^{2}$. The bin widths are $\delta u=\delta \gamma =0.5$ in units of their rms.}
  \label{fig:8m}
\end{figure}

Recent experimental data have been taken behind a grid put in the test section of the wind tunnel  in Modane, see Fig.1 of Ref. \cite{mickael}. We got some of them in order to see if the location of the hot-wires has an effect on the detection of eventual singularities in the flow.   Among the  several files of velocity recorded by hot wires that we have investigated,  we present here two of them recorded  at  two  different locations.  In  the first record the hot-wire was placed far from the grid (at $23m$ behind it),  where the turbulence is supposed to have relaxed to an isotropic and homogeneous state. The second record was taped  closer to  the grid (at 8m behind it). In both cases the mean velocity  is twice larger than in the previous study, the  Reynolds number is  about five times smaller, $Re_{\lambda}\simeq 500$, and the sampling frequency is $250$ KHz (ten times larger than in the ancient data)\footnote{When studying  the recent data  with sampling frequency equal to $250$ KHz, we  have observed  white areas in the  join probability $P(u,\gamma)\delta u\delta \gamma$ when the  bin widths were smaller than a certain value.  For that reason we chose bin widths equal to $0.5$   in units of rms to calculate the conditional momenta. The white areas (without any points) show up as  quasi-parallel rows, regularly arranged in the plane $(u,\gamma)$, even in the domain where the number of points is maximum.  We attribute this effect to the fact that the sampling time is probably too small, perhaps $3$ or $5$ times shorter than the response time of the hot wire. A way to suppress these white zones is to  filter the raw data. We used also this technique, which reduces the amplitude of the fluctuations and  smoothen the signals, and checked that  it gives  results (not shown here) in agreement with  those presented in  Sec.\ref{sec:modane2}.}. The record duration is $10$ and $13$ min. respectively, that gives  files with $150$ and $200$ millions of points.   The result of our analysis of conditional moments given the acceleration, are presented in Figs.~\ref{fig:23m} and \ref{fig:8m}.
 
  In the first case,  far from the grid,
  the   curves  $\left<u^{3}\right>_{\gamma}$  versus $\gamma$  in Fig.\ref{fig:23m}-(a) and $\left<u^{3}\right>_{\gamma}/ \gamma$  in   Fig.\ref{fig:23m}-(b),  display  oscillations not in agreement with self-similar scalings. One may possibly observe that  a linear relation  $\left<u^{3}\right>_{\gamma}\propto \gamma$ occurs for very large acceleration values, but in this domain the number of points is very small,  the total number points  corresponding to the  linear domain being less than $0.1$ per thousand. 
 The slope of the curve is  very small, it would correspond to a small circulation and to a local Reynolds number of order $Re_{\rm local}\approx 0.16$. This value  smaller than unity  points towards events occurring in the dissipative domain only and to rare singular events where inertia is dominant.
 
 Our interpretation of  the small circulation deduced from this record, is that the hot-wire is located in a zone of  decaying turbulence, where few organized structures with a large circulation  have survived, structures that could become singularities. To  assert this we  have investigated other data files recorded closer to the grid (at  8m from the   grid).  These data display some events with huge acceleration, the largest one being of order $300$ times the rms (see below Fig.~\ref{fig:burst})! 
 The result of the statistical study is shown in Figs.~\ref{fig:8m}. It displays a significant domain where $\left<u^{3}\right>_{\gamma}$ is proportional to $\gamma$, moreover the  slope is about  $30$ times larger than far from the grid, that gives a local Reynolds number larger than unity, $Re_{\gamma} \simeq 5$.  Therefore the  proximity of the grid clearly helps the formation of singular structures (if any).   For comparison with the ancient data taken in the  return vein of the wind tunnel, Y. Gagne informed us that the hot wire was placed on the axis in a zone of chunk turbulence. In this place the large scale flow results from several  contributions, a strongly diverging main flow (the diameter of the vein being 2.4 the one of the test domain), plus additional incoming cold wakes from far entrance. Moreover a large protective grid is  placed in the return vein, against  possible flying objects.  The good fit of these data with Leray's scalings allows us  to conjecture that those additive contributions  could help for the formation of  singularities.
 
\subsection {Asymmetry of large fluctuations}
\begin{figure}
\centerline{
 \includegraphics[width=2.5in]{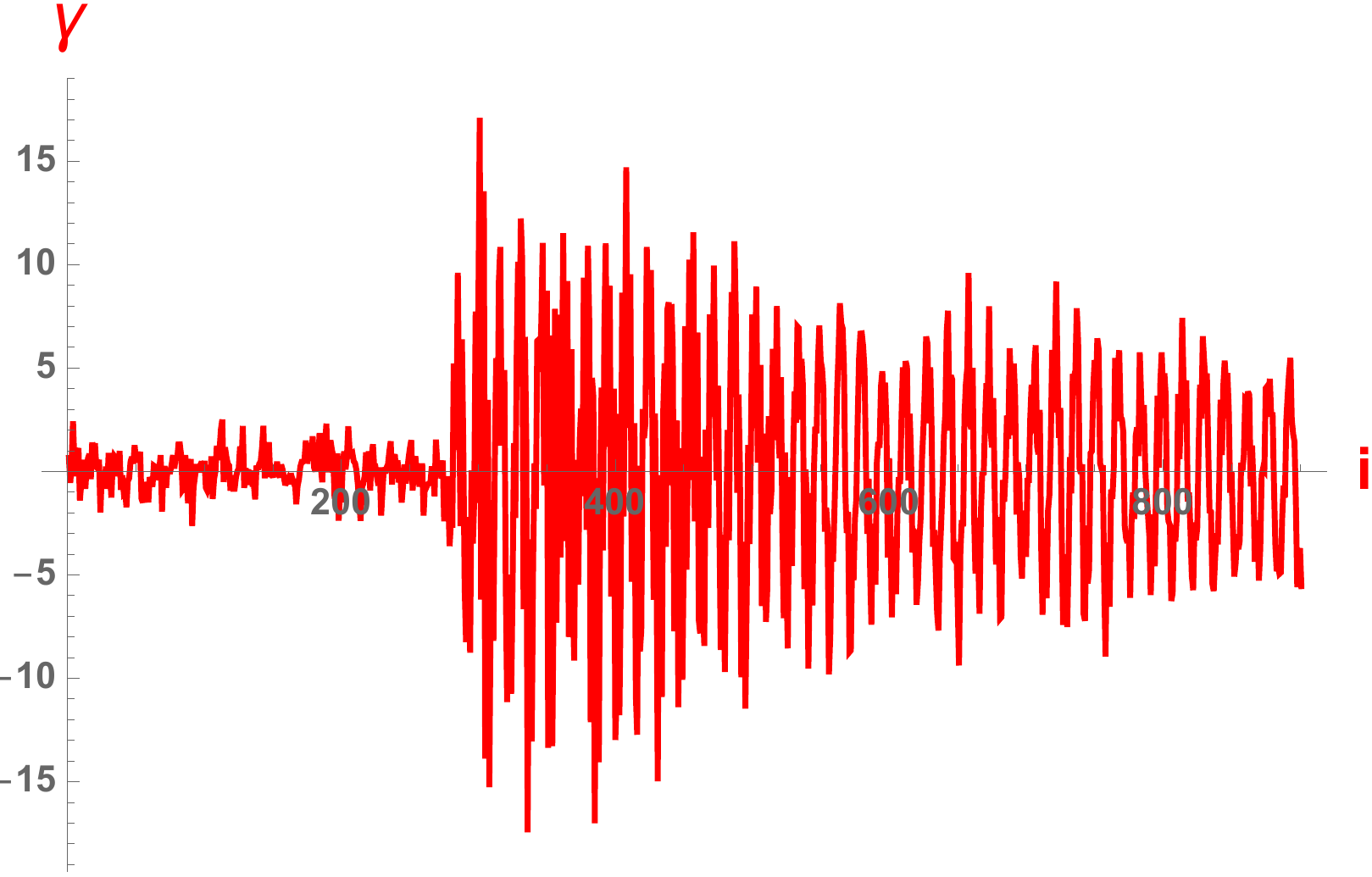}
 }
 \centerline{
  \includegraphics[width=2.5 in]{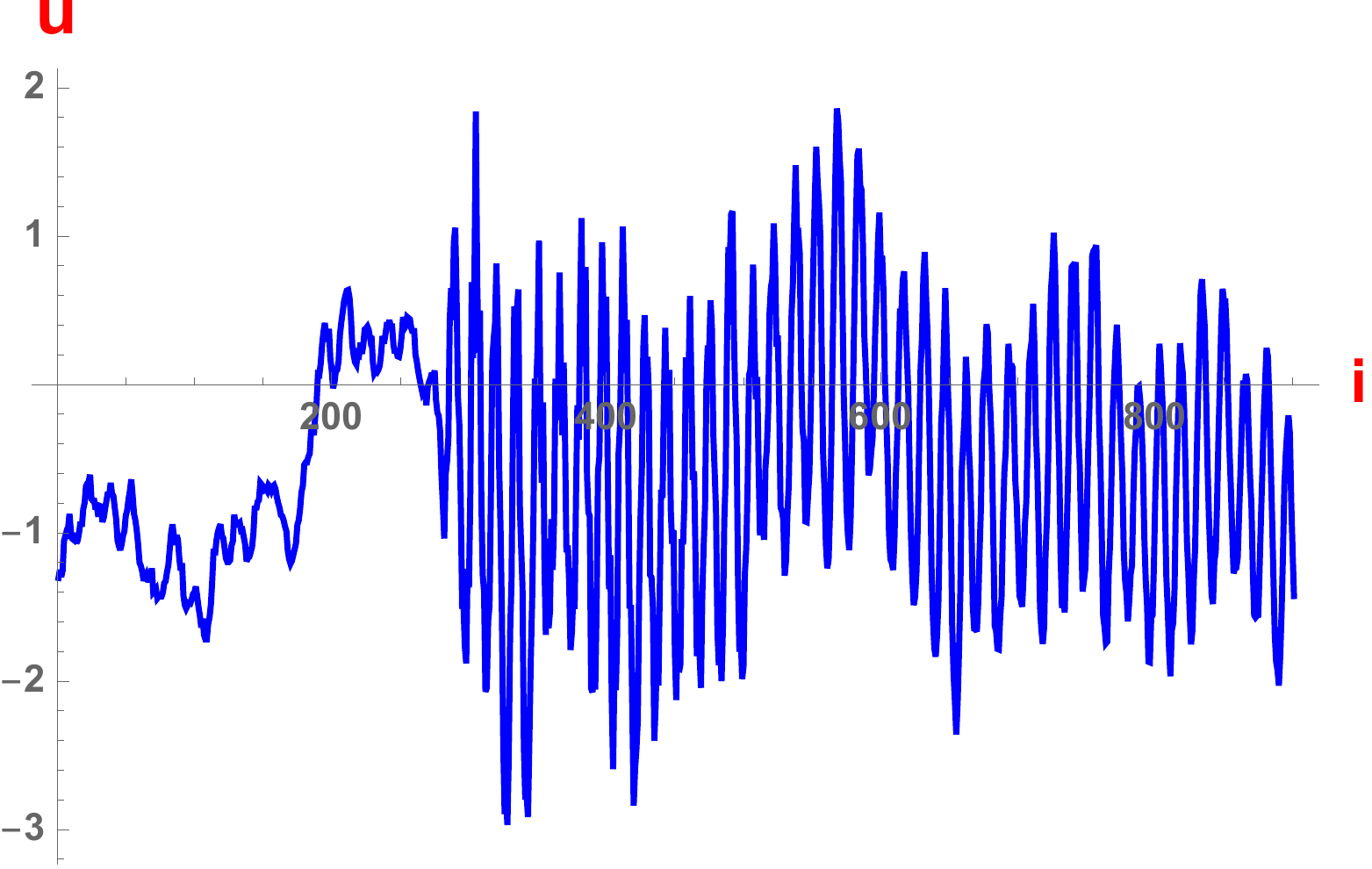} 
 }
 \caption{ Asymmetric burst of acceleration and velocity.}
\label{fig:burst}
\end{figure}

This section is to point out that the records of very large bursts  are
strongly asymmetric, meaning that the observed rise of the amplitude of
velocity and acceleration is very quick and barely observable at our time resolution whereas the
decay of large fluctuations is quite long and involves many
oscillations. The maximum amplitude of acceleration observed in  Modane comes from the recent
records of $v(t)$ made  at high sampling frequency close to the grid, where  the data show a strongly asymmetric burst, with acceleration as large as  $10^{7}\;m/s^{2}$,  ($280\sigma_{\gamma}$).  Similar bursts  are also observed  for lower amplitude peaks, of order $10^{5}-10^{6}\;m/s^{2}$, as shown  in Fig.\ref{fig:burst} where the maximum is about $15 \sigma_{\gamma}$. 

 \begin{figure}
\centerline{
 \includegraphics[height=1.5 in]{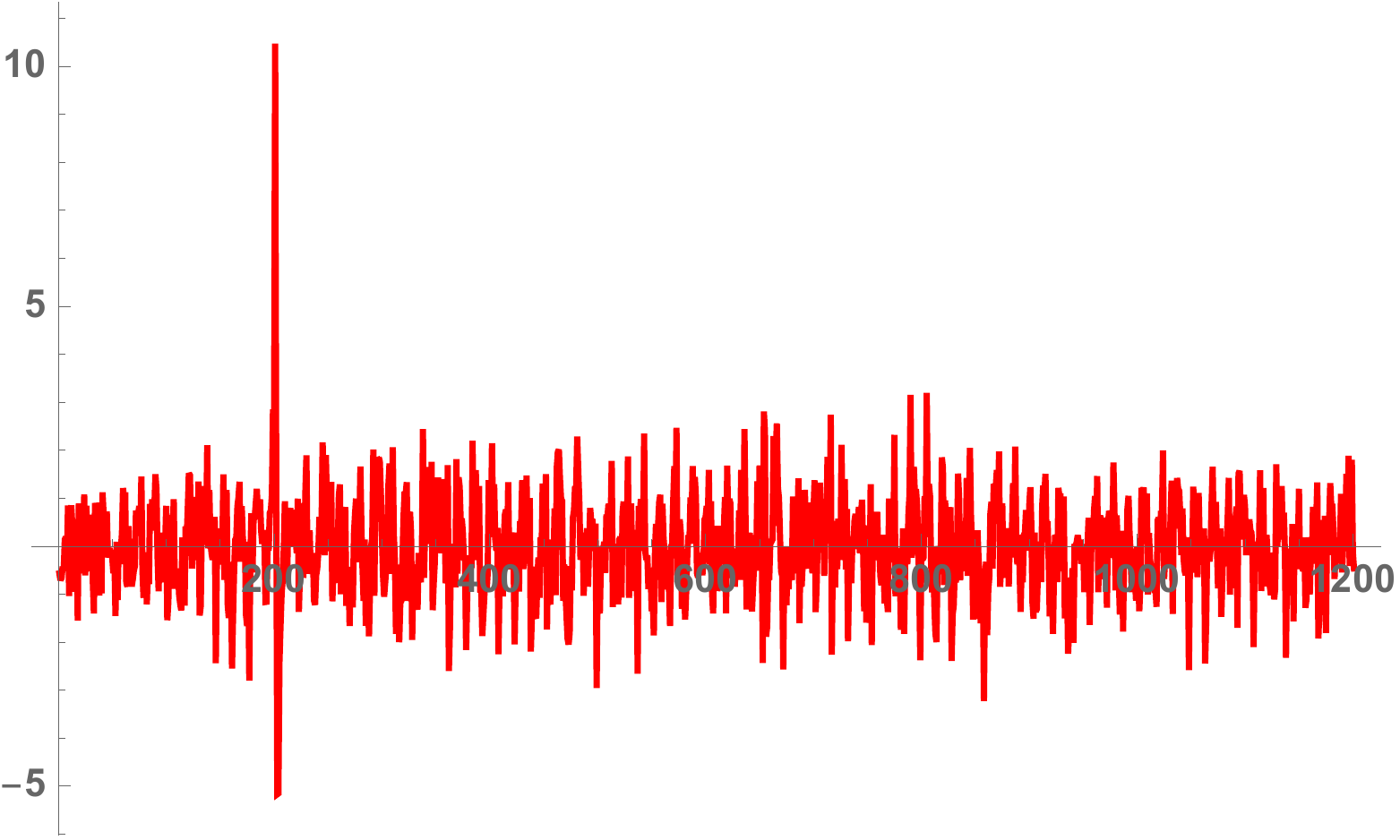}
  \includegraphics[height=1.5 in]{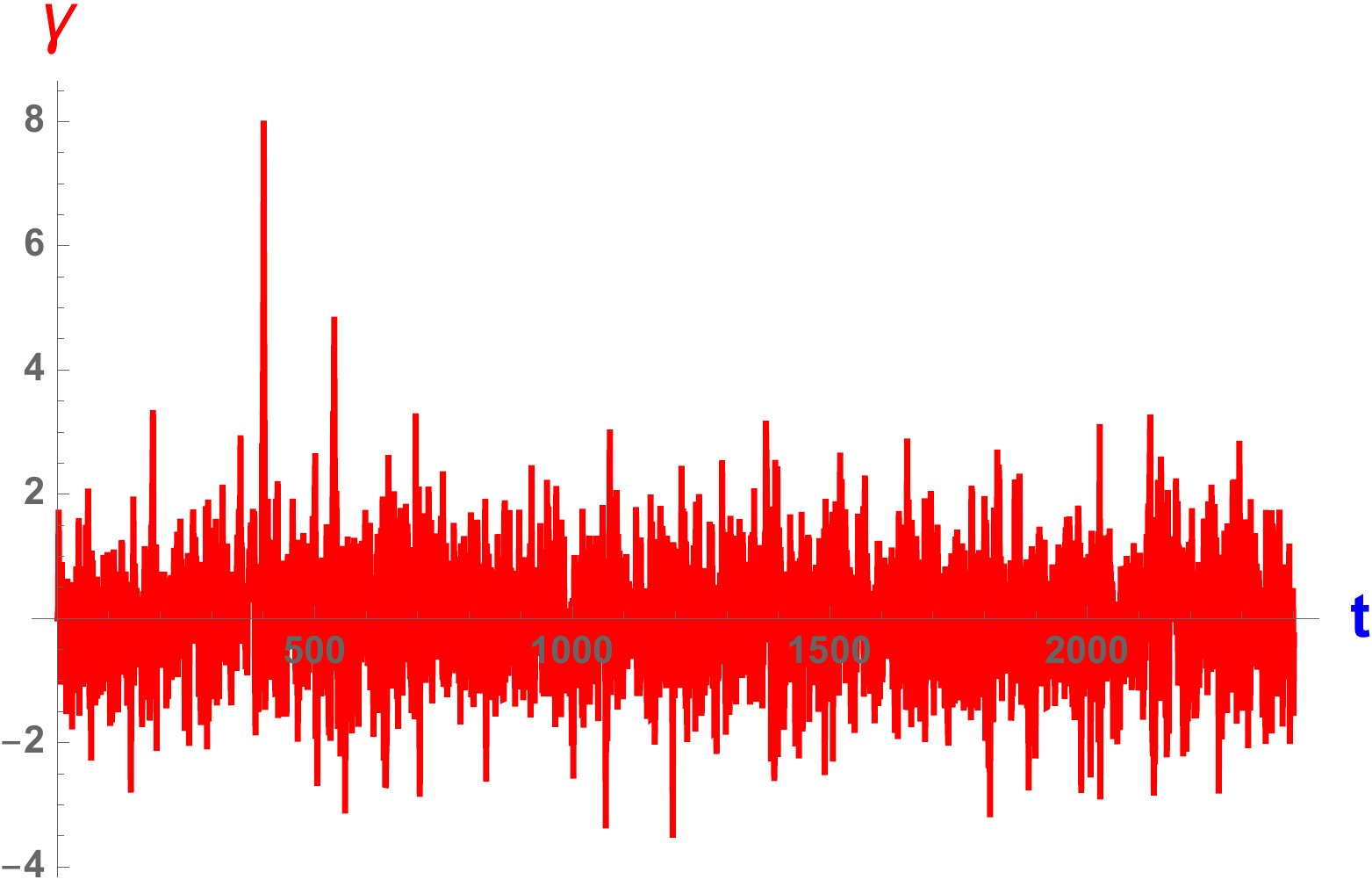} 
 }
 \centerline{
  \includegraphics[height=1.5 in]{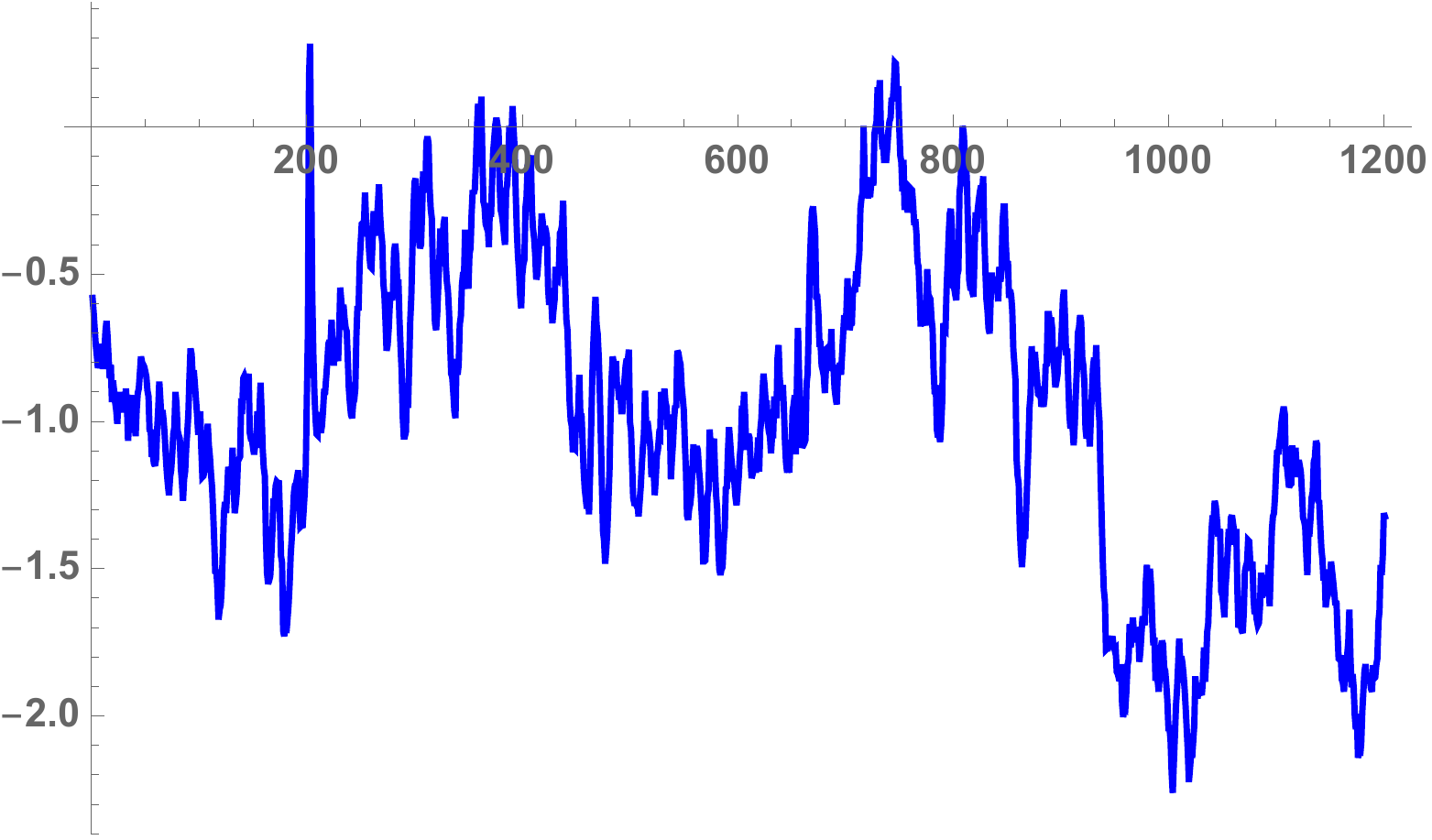}
  \includegraphics[height=1.5 in]{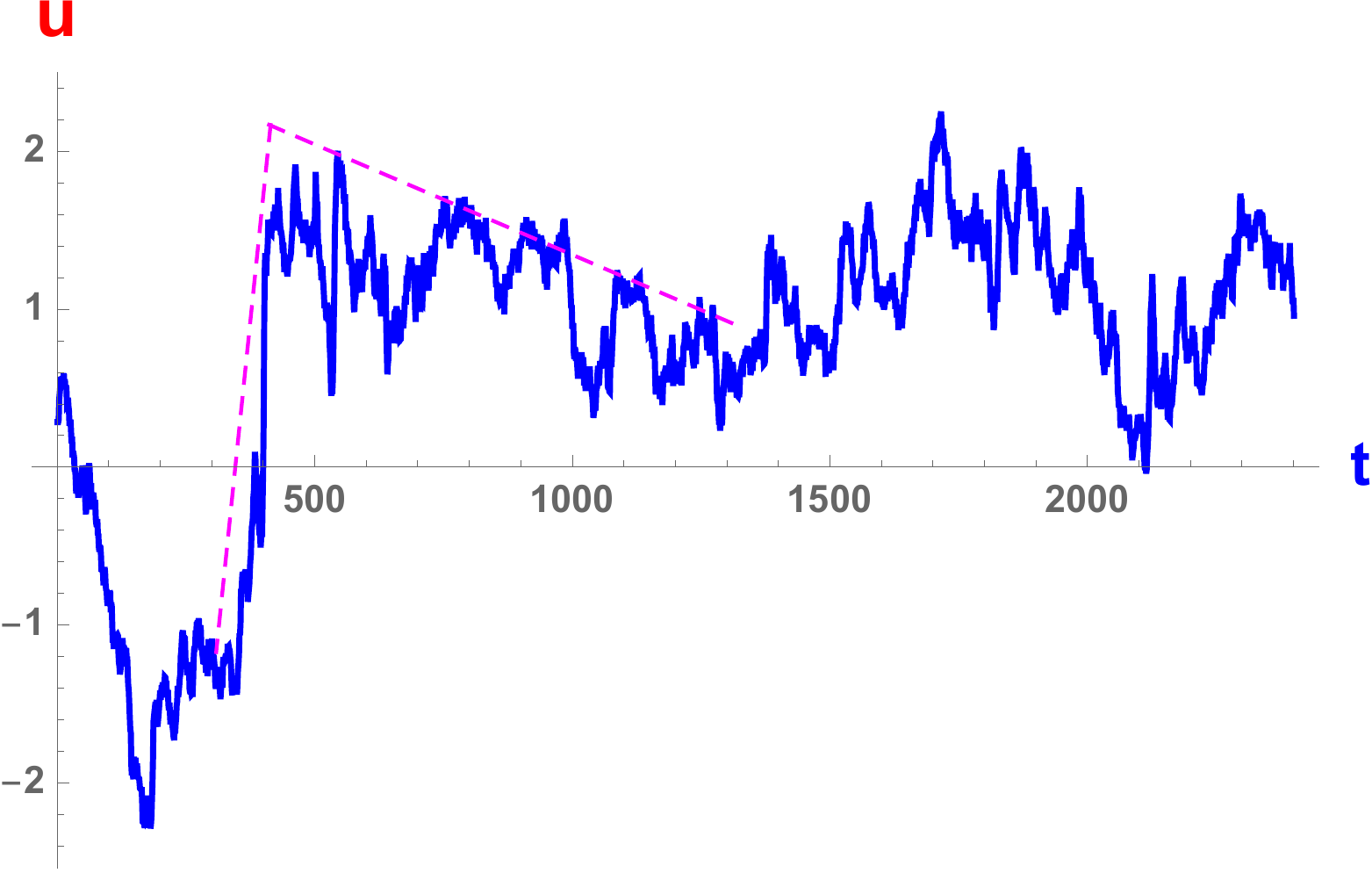}  
   }
  \caption{ left: Isolated peak of acceleration and velocity. Right: Successive peaks  of acceleration, the asymmetry of rising and decay time is suggested by the dashed line on the velocity plot. The acceleration (red curves) and velocity fluctuation (blue curves) are plotted in units of their respective standard deviation. Time is in units of the sampling time.}
\label{fig:peak}
\end{figure}

Such an asymmetry is expected, as explained on general grounds (irreversibility with respect to time)  \cite{modane}, but this general property does not help much to explain such spectacular recording. We believe now that the asymmetry of the time signal  $u(t)$ close to a peak could provide one argument in favor of the existence of finite time
singularities (in Euler equations) to explain intermittency in high Reynolds number flows.
The idea for explaining the striking asymmetry of Fig.~\ref{fig:burst} is based on the remark that the growth of the
bursts is described as an incipient singularity of a solution of the Euler
inviscid equation. Turning on the viscosity, as we show in section \ref{sec:smallvisc} of
this paper, makes drift the singular solution toward lower and lower
amplitudes as it gets closer and closer to the time of blow-up. At the end
of this smoothening the fluid motion becomes ruled by the NS equation at
finite Reynolds number. We suggest that, when this happens, the
fluctuation decays far more slowly than it has grown because of the
decay of the non linear part of the dynamics. Therefore the typical
time scale becomes much longer, as observed, as well as the magnitude of
the acceleration. The rather complex pattern of time dependence should be the result of
oscillations linked to the fact that, even though the Reynolds number is
not infinite, the relaxation is still oscillatory because of the effects of
the finite nonlinearity of the fluid equations.

We stress that some large peaks  show up as isolated ones,  as  in  left Figs.\ref{fig:peak}, or  else as successive peaks of acceleration associated to an asymmetric velocity signal, as shown in the right figures.  It is important to notice that the two regimes (quick
growth and slow decay) could well be even more different of each other than in the figures,
due to the finite time resolution of the measuring device.

\section{ Summary and conclusion}
\label{summ} 
We discussed the existence of singular solutions of the inviscid and incompressible fluid equations and how this is related to experimental data. Assuming that the Navier-Stokes equations have  no truly singular relevant solution whereas the Euler equations have such singular solutions, we derive an equation for the decay of the singular solution of the Euler -Leray singularity under the effect of viscosity. Beyond the theoretical analysis we compare a prediction of the self-similar dynamics with experimental records. It has been known since Batchelor and Townsend that turbulent flows generate large  and short lived derivatives of the velocity fluctuations. The relationship we uncover between large accelerations and large velocities  agrees  with our  explanation of this observed intermittency as due to singularities of solutions of Euler equations. 

On a wider point of view, this also shows that, perhaps, more is to be expected  in the understanding of turbulence from solutions of the time dependent fluid equations, including possible effect of  a small but non vanishing viscosity, something which is not so surprising after all!

Using scalings  deduced from Leray singular solutions, we have shown that experimental data  recorded in the wind tunnel of  Modane are compatible with such sparse solutions which could well be not so rare in the case of high Reynolds number and for sufficient injected circulation.
Because intermittency can be seen as a strong deviation from K41 scaling law, it is not  new to find experimental  data  which deviate  from K41 scalings. We point out that the well-known relation $u_{r}=(\epsilon r)^{1/3}$  for the velocity fluctuations between two points  separated by a distance $r$
 results from the hypothesis that the dissipation per unit mass, $\epsilon$, is uniform in space and time.
If the exponent of $r$  is less than
$1/3$,  and if the power  dissipated par unit mass is  still assumed to be uniform in space, the dissipation should diverge.  In Modane we have found a  negative exponent, namely a relation fitting the scaling $u_{r} \sim r^{-x}$ with $x$ of order unity as predicted by (\ref{eq:moment3}). Therefore  finite dissipation (on average) and  
 an exponent of $r$  less than 1/3 , as found in Modane, can
be explained by a sparse (with
zero measure) support in space-time of dissipation events. Such a scenario is well
explained by the random occurrence in space-time of singularities of the
Leray type. Somehow this connects well the statistical properties of a turbulent flow with the solution of the fluid equations.

\section*{Acknowledgements}
Christophe Josserand, B\'erang\`{e}re Dubrulle, Sergio Rica and St\'ephane Popinet are greatly acknowledged for stimulating and fruitful discussions.  We also acknowledge Yves Gagne for giving us precisions about the data of  the $90$s, and Mickael Bourgoin who provide us the data files  recorded  during the recent experiment  carried out in the S1MA wind-tunnel from ONERA in Modane in the context of the ESWIRP European project.

\end{document}